\documentclass[aps,prd,preprintnumbers,groupedaddress,nofootinbib,amssymb,eqsecnum,notitlepage]{revtex4-1}
\usepackage[dvipdfmx]{graphicx}
\usepackage{color}
\usepackage{here}
\usepackage{amsmath,amsthm,amssymb}
\usepackage{bm}
\allowdisplaybreaks[1]

\usepackage{amsfonts}
\usepackage{dcolumn}
\usepackage[dvipdfmx]{hyperref}

\begin{document}
\newcommand{\newc}{\newcommand}
\newc{\tif}{\tilde{f}}
\newc{\tih}{\tilde{h}}
\newc{\tip}{\tilde{\phi}}
\newc{\tiA}{\tilde{A}}

\newcommand{\ben}{\begin{eqnarray}}
\newcommand{\een}{\end{eqnarray}}
\newc{\be}{\begin{equation}}
\newc{\ee}{\end{equation}}
\newc{\ba}{\begin{eqnarray}}
\newc{\ea}{\end{eqnarray}}
\newc{\bea}{\begin{eqnarray*}}
\newc{\eea}{\end{eqnarray*}}
\newc{\D}{\partial}
\newc{\ie}{{\it i.e.} }
\newc{\eg}{{\it e.g.} }
\newc{\etc}{{\it etc.} }
\newc{\etal}{{\it et al.}}
\newcommand{\nn}{\nonumber}
\newc{\ra}{\rightarrow}
\newc{\lra}{\leftrightarrow}
\newc{\lsim}{\buildrel{<}\over{\sim}}
\newc{\gsim}{\buildrel{>}\over{\sim}}
\newc{\aP}{\alpha_{\rm P}}
\newc{\dphi}{\delta\phi}
\newc{\da}{\delta A}
\newc{\tp}{\dot{\phi}}

\title{Dark energy in scalar-vector-tensor theories}

\author{Ryotaro Kase and Shinji Tsujikawa}

\affiliation{
Department of Physics, Faculty of Science, 
Tokyo University of Science, 1-3, Kagurazaka,
Shinjuku-ku, Tokyo 162-8601, Japan}

\date{\today}

\begin{abstract}

The scalar-vector-tensor theories with second-order equations 
of motion can accommodate both 
Horndeski and generalized Proca theories as specific cases. 
In the presence of a perfect fluid, we study the cosmology 
in such a most general scheme of scalar-vector-tensor theories 
with parity invariance by paying particular attention to the application to dark energy.
We obtain a closed-form expression of the background 
equations of motion by using coefficients appearing in 
the second-order action of scalar perturbations. 
We also derive conditions for the absence of ghost 
and Laplacian instabilities of tensor and vector perturbations and show that 
the existence of matter does not substantially modify the stabilities of 
dynamical degrees of freedom in the small-scale limit. 
On the other hand, the sound speed of scalar perturbations is affected 
by the presence of matter.
Employing the quasi-static approximation for scalar perturbations 
deep inside the sound horizon, we derive analytic expressions of 
Newtonian and weak lensing  gravitational potentials 
as well as two scalar perturbations
arising from the scalar and vector fields. 
We apply our general framework to dark energy 
theories with the tensor propagation speed equivalent to 
the speed of light  
and show that the observables associated with the growth 
of matter perturbations and weak lensing potentials 
are generally affected by intrinsic vector modes and by interactions between scalar and vector fields.

\end{abstract}


\maketitle

\section{Introduction}
\label{introsec}

Since the first discovery of late-time cosmic acceleration 
in 1998 \cite{SN1,SN2}, the origin of dark energy has not been identified yet. The cosmological constant is overall consistent 
with the current observational data, but there is still a discrepancy 
between the values of today's Hubble constant $H_0$ constrained from the Cosmic Microwave Background (CMB) data \cite{Planck2015,Planck2018} and from local 
measurements at low redshifts \cite{RiessH0}. 
In dynamical models of dark energy whose equation of state 
$w_{\rm DE}$ varies in time, there are some possibilities 
for relaxing this tension of $H_0$ \cite{Gong1,Valentino,Gong2}.

The minimally coupled scalar fields like quintessence \cite{quin} 
and k-essence \cite{kes} predict the dynamical dark energy 
equation of state $w_{\rm DE}$ larger than $-1$. 
The observational data of CMB combined with 
the data of baryon acoustic oscillations and type Ia 
supernovae have allowed not only $w_{\rm DE}$ larger than $-1$ but also the region $w_{\rm DE}<-1$ \cite{CDT,CMB,Jean}.  
The latter region can be realized by the existence of 
a scalar or vector field coupled to gravity \cite{review1,review2}. 
Indeed, it was shown that dark energy 
models based on extended scalar Galileons \cite{exga} 
or generalized Proca theories \cite{GPcosmo} 
can reduce the tension of the $\Lambda$-cold-dark-matter model 
mentioned above \cite{DT12,DHT17}.

For a scalar field $\phi$ coupled to gravity, Horndeski 
theories \cite{Horndeski} are the most general scalar-tensor theories 
with second-order equations of motion \cite{Horn1,Horn2,Horn3}, 
which accommodate (extended) Galileons \cite{Nicolis,Galileon,exga} 
as specific cases. 
The application of Horndeski theories to dark energy has been extensively 
carried out in the literature  \cite{GSami,DT10,Gama,DKT,CCPS,Amendola12,Miguel,Bellini}. 
The modification of gravity from General Relativity (GR) gives rise to the speed of gravitational waves $c_t$ 
which is not necessarily equivalent to that of light $c$. 
On the other hand, the detection of the gravitational-wave event 
GW170817 \cite{GW170817} from a neutron star merger together 
with the gamma-ray burst GRB 170817A \cite{Goldstein} constrained 
$c_t$ to be very close to $c$ \cite{Abbott}.
If we impose $c_t=c$ exactly and do not allow the tuning between 
functions, the Lagrangian of Horndeski theories needs to be 
of the form ${\cal L}=G_2(\phi,X_1)+G_3(\phi,X_1) \square \phi
+G_4(\phi)R$ \cite{Lucas,GW1,GW2,GW3,GW4,GW5,GW6,KT18}, 
where $G_2, G_3$ are functions of $\phi$ 
and $X_1=-\partial_{\mu}\phi \partial^{\mu}\phi/2$, 
$G_4$ is a function of $\phi$, and $R$ is the Ricci scalar. 
The Brans-Dicke theory \cite{Brans}, $f(R)$ gravity \cite{fR}, 
and cubic Galileons \cite{Nicolis,Galileon} belong to 
the theories with $c_t=c$.

For a vector field $A_{\mu}$ coupled to gravity with broken $U(1)$ gauge 
symmetry, generalized Proca theories \cite{Heisenberg,Tasinato,Allys,Jimenez} 
are the most general 
vector-tensor theories with second-order equations of motion 
(see Refs.~\cite{Gripaios,Deffayet} for earlier works). 
If we apply generalized Proca theories to cosmology, 
the existence of vector derivative and nonminimal 
couplings to gravity can lead to the late-time cosmic acceleration with a temporal vector 
component $A_0$ \cite{GPcosmo}. 
Similar mechanisms of the Universe acceleration were 
also advocated in Refs.~\cite{VTearly}. 
Dark energy models in generalized Proca theories can leave several 
interesting observational signatures such as the constant 
equation of state $w_{\rm DE}<-1$ in the matter era \cite{GPcosmo} 
and the possibility for realizing the cosmic growth rate slower 
than that in GR \cite{Geff16}. 
Moreover, the propagation of fifth forces in local regions of 
the universe can be suppressed under the operation of the 
Vainshtein mechanism \cite{GPsc}.
Imposing the constraint $c_t=c$, the dependence 
of $X_3=-A_{\mu}A^{\mu}/2$ in quartic and quintic couplings 
$G_4(X_3)$ and $G_5(X_3)$ are not allowed, while all the 
other interactions including intrinsic vector modes are possible.
This restricts the cosmic growth rate in the range larger than 
that in GR \cite{GW5}, but the evolution of $w_{\rm DE}$ mentioned 
above is still possible.

Horndeski and generalized Proca theories can be united in the form 
of scalar-vector-tensor (SVT) theories with second-order equations 
of motion \cite{Heisenberg18}.
In SVT theories with $U(1)$ gauge symmetry, the longitudinal component 
of the vector field $A_{\mu}$ does not propagate. 
Then, the dynamical degrees of freedom (DOFs) are the scalar field 
$\phi$, two transverse vector modes associated with $A_{\mu}$, 
and two tensor polarizations arising from the gravity sector. 
The Lagrangian of $U(1)$-invariant SVT theories was constructed in 
Ref.~\cite{Heisenberg18}, which was recently applied to the study of 
hairy black hole solutions and their stabilities \cite{HT18,Cha,Pedro,HKTodd}.
The SVT theories with broken $U(1)$ gauge symmetry give rise to 
the additional longitudinal propagation of $A_{\mu}$, so there are 
six propagating DOFs in total. 
Moreover, the temporal vector component $A_0$ contributes to the 
background cosmological dynamics besides 
the scalar field $\phi$. If we apply such theories 
to inflation, for example, the standard single-field dynamics 
driven by $\phi$ is modified by the auxiliary field $A_0$ \cite{HKT18}.

In the presence of new interactions arising in SVT theories 
with broken $U(1)$ gauge symmetry, the authors of Ref.~\cite{HKT18} 
derived the second-order actions of tensor, vector, and scalar perturbations 
on the flat Friedmann-Lema\^{i}tre-Robertson-Walker (FLRW) background 
to elucidate conditions for the absence of ghost and Laplacian instabilities 
in the small-scale limit.
While these results can be directly applied to the inflationary epoch 
in which the contribution of additional matter 
to the cosmological dynamics is neglected, this is not the case
for the dynamics of late-time cosmic acceleration during which the energy densities 
of dark matter and baryons cannot be ignored relative to those of dark energy. 
In this paper, we consider SVT theories with broken $U(1)$ gauge symmetry 
by implementing a perfect fluid matter described by 
a Schutz-Sorkin action \cite{Sorkin,DGS}.
This gives rise to the additional scalar propagation, 
so there are three dynamical scalar DOFs in addition to two vector 
and two tensor propagating DOFs.

Besides the SVT interactions ${\cal S}_{{\rm SVT}}$ studied
in Ref.~\cite{HKT18}, we take into account the Horndeski action 
${\cal S}_{{\rm ST}}$ to accomodate full parity-invariant 
SVT theories with second-order equations of motion. 
We derive the background equations 
of motion in a closed form and then expand the full SVT action 
in the presence of matter up to quadratic order in tensor, vector, and scalar perturbations.
To test for SVT theories with the observations of large-scale 
structures and weak lensing, we analytically compute two gauge-invariant 
gravitational potentials as well as scalar perturbations arising from $\phi$ and $A_{\mu}$ by employing the so-called quasi-static 
approximation \cite{Boi,review1,DKT} 
for the modes deep inside the sound horizon.

Our general analysis encompasses both Horndeski and generalized Proca theories as specific cases. 
The extension of Horndeski and generalized Proca theories to 
the domain of SVT theories opens up a new window 
for the dynamics of dark energy and the cosmic growth history.
At the background level, the interaction between scalar 
and vector fields affects the evolution of the dark energy 
equation of state. The growth of cosmological perturbations 
can be generally modified not only by scalar-vector 
interactions but also by intrinsic vector modes. 
We also study the case in which the condition $c_t=c$ is 
imposed and discuss new features arising in Newtonian 
and weak lensing gravitational potentials.

Our paper is organized as follows.
In Sec.~\ref{modelsec}, we present the action of most general 
SVT theories with second-order equations 
of motion in the presence of matter.
In Sec.~\ref{backsec}, we express the background equations of 
motion in a compact form by using coefficients arising in the 
second-order action of scalar perturbations.
In Sec.~\ref{TVsec}, we derive no-ghost conditions as well as 
the propagation speeds of tensor and vector perturbations in 
the small-scale limit. 
In Sec.~\ref{scasec}, we obtain the full scalar perturbation equations 
and clarify conditions for the absence of scalar ghost and 
Laplacian instabilities. 
In Sec.~\ref{Geffsec}, the analytic expressions of Newtonian and 
weak lensing gravitational potentials are derived under the quasi-static 
approximation to confront SVT theories with observations 
associated with the cosmic growth history.
In Sec.~\ref{conmodelsec}, we apply our general formulas of 
gravitational potentials for the SVT theories in which the speed 
of gravity is equivalent to $c$.  
Sec.~\ref{concludesec} is devoted to conclusions. 
In what follows, we use the natural unit $c=1$.

\section{SVT theories with broken $U(1)$ gauge invariance}
\label{modelsec}

The SVT theories with broken $U(1)$ gauge 
symmetry \cite{Heisenberg18} contain 
a scalar field $\phi$ and a vector field $A_{\mu}$ coupled to gravity. 
To describe kinetic terms of $\phi$ and $A_{\mu}$ and their interactions, we define 
\be
X_1=-\frac{1}{2} \nabla_{\mu} \phi \nabla^{\mu} \phi\,,
\qquad
X_2=-\frac{1}{2} A^{\mu} \nabla_{\mu} \phi \,,
\qquad 
X_3=-\frac{1}{2} A_{\mu} A^{\mu}\,,
\label{X123}
\ee
where $\nabla_{\mu}$ represents a covariant derivative operator. 
We introduce a symmetric tensor $S_{\mu\nu}$ 
constructed from $A_{\mu}$, as 
\be
S_{\mu \nu}=\nabla_{\mu}A_{\nu}
+\nabla_{\nu}A_{\mu}\,,
\ee
together with the antisymmetric field strength tensor 
$F_{\mu \nu}$ and its dual $\tilde{F}_{\mu \nu}$, as 
\be
F_{\mu\nu}=\nabla_\mu A_\nu-\nabla_\nu A_\mu\,,
\qquad 
\tilde{F}^{\mu\nu}=\frac{1}{2}
\mathcal{E}^{\mu\nu\alpha\beta}F_{\alpha\beta}\,,
\ee
where $\mathcal{E}^{\mu\nu\alpha\beta}$ is the anti-symmetric Levi-Civita tensor 
obeying the normalization $\mathcal{E}^{\mu\nu\alpha\beta}
\mathcal{E}_{\mu\nu\alpha\beta}=-4!$. 
We define several Lorentz-invariant quantities associated with 
intrinsic vector modes, as
\be
F=-\frac{1}{4}F_{\mu\nu}F^{\mu\nu}\,,\qquad
Y_1=\nabla_\mu \phi \nabla_\nu \phi\,F^{\mu\alpha}F^\nu{}_\alpha\,,
\qquad 
Y_2=\nabla_\mu\phi\, A_\nu F^{\mu\alpha}F^\nu{}_\alpha\,, 
\qquad
Y_3=A_\mu A_\nu F^{\mu\alpha}F^\nu{}_\alpha\,,
\label{FY123}
\ee
which vanish by taking the scalar limit 
$A_{\mu} \to \nabla_{\mu} \pi$.

We consider the SVT interactions described by the action \cite{Heisenberg18}:
\be
\mathcal{S}_{\rm SVT}=
\int d^4x \sqrt{-g}\,\sum_{n=2}^6
\mathcal{L}_{{\rm SVT}}^{(n)}\,,
\label{action}
\ee
with the Lagrangians
\ba
\mathcal{L}_{{\rm SVT}}^{(2)} &=&
f_2(\phi,X_1,X_2,X_3,F,Y_1,Y_2,Y_3)\,, \nonumber\\
\mathcal{L}_{{\rm SVT}}^{(3)}  &=&
f_{3}(\phi,X_3)g^{\mu\nu}S_{\mu\nu}
+\tilde{f}_{3}(\phi,X_3)A^{\mu}A^{\nu} S_{\mu\nu}\,,\nonumber\\
\mathcal{L}_{{\rm SVT}}^{(4)} & = & 
f_{4}(\phi,X_3)R+f_{4,X_3}(\phi,X_3) \left[ 
(\nabla_\mu A^\mu)^2-\nabla_\mu A_\nu \nabla^\nu A^\mu \right]\,, \nonumber\\
\mathcal{L}_{{\rm SVT}}^{(5)} & = & 
f_5(\phi,X_3)G^{\mu\nu} \nabla_{\mu}A_{\nu}
-\frac{1}{6}f_{5,X_3}(\phi,X_3) 
\left[ (\nabla_{\mu} A^{\mu})^3
-3\nabla_{\mu} A^{\mu}
\nabla_{\rho}A_{\sigma} \nabla^{\sigma}A^{\rho}
+2\nabla_{\rho}A_{\sigma} \nabla^{\gamma}
A^{\rho} \nabla^{\sigma}A_{\gamma}\right] \nonumber\\
&+&\mathcal{M}_5^{\mu\nu}\nabla_\mu \nabla_\nu\phi
+\mathcal{N}_5^{\mu\nu}S_{\mu\nu}\,, 
\nonumber\\
\mathcal{L}_{{\rm SVT}}^{(6)} & = &
f_6(\phi,X_1)L^{\mu\nu\alpha\beta}F_{\mu\nu}F_{\alpha\beta}
+\mathcal{M}_6^{\mu\nu\alpha\beta}\nabla_\mu\nabla_\alpha \phi\nabla_\nu
\nabla_\beta\phi+\tilde{f}_6(\phi,X_3)L^{\mu\nu\alpha\beta}F_{\mu\nu}F_{\alpha\beta}+ \mathcal{N}_6^{\mu\nu\alpha\beta}S_{\mu\alpha}S_{\nu\beta}\,,
\label{LagSVT}
\ea
where $g$ is a determinant of the metric tensor $g_{\mu\nu}$, 
$R$ and $G^{\mu \nu}$ are the Ricci scalar and the 
Einstein tensor, respectively, and $L^{\mu\nu\alpha\beta}$ 
is the double dual Riemann tensor defined by 
\be
L^{\mu\nu\alpha\beta}=\frac{1}{4}
\mathcal{E}^{\mu\nu\rho\sigma}
\mathcal{E}^{\alpha\beta\gamma\delta} 
R_{\rho\sigma\gamma\delta}\,,
\ee
where $R_{\rho\sigma\gamma\delta}$ is the Riemann tensor. 
The function $f_2$ depends on $\phi, X_i, F, Y_i$, where the 
subscript represents $i=1,2,3$.  
The functions $f_3, \tilde{f}_3, f_4, f_5, \tilde{f}_6$ are dependent on $\phi$ and $X_3$, while $f_6$ 
is a function of $\phi$ and $X_1$. 
For partial derivatives with respect to $\phi, X_i, F, Y_i$,
we use the notations like 
$f_{4,X_3} \equiv \partial f_4/\partial X_3$.
 
The 2-rank tensors $\mathcal{M}^{\mu\nu}_5$ and 
$\mathcal{N}^{\mu\nu}_5$ in ${\cal L}_{\rm SVT}^{(5)}$, 
which are associated with intrinsic vector modes, are 
given by 
\be
\mathcal{M}^{\mu\nu}_5
=\mathcal{G}_{\rho\sigma}^{h_{5}} 
\tilde{F}^{\mu\rho}\tilde{F}^{\nu\sigma}\,,\qquad 
\mathcal{N}^{\mu\nu}_5
=\mathcal{G}_{\rho\sigma}^{\tilde{h}_{5}} 
\tilde{F}^{\mu\rho}\tilde{F}^{\nu\sigma}\,,
\ee
with
\ba
\mathcal{G}_{\rho\sigma}^{h_5} 
&=& 
h_{51}(\phi,X_i)g_{\rho\sigma}+h_{52}(\phi,X_i)
\nabla_\rho \phi \nabla_\sigma \phi
+h_{53}(\phi,X_i)A_\rho A_\sigma
+h_{54}(\phi,X_i)A_\rho \nabla_\sigma \phi\,,\\
\mathcal{G}_{\rho\sigma}^{\tilde{h}_5} 
&=& 
\tilde{h}_{51}(\phi,X_i)g_{\rho\sigma}
+\tilde{h}_{52}(\phi,X_i)
\nabla_\rho \phi \nabla_\sigma \phi
+\tilde{h}_{53}(\phi,X_i)A_\rho A_\sigma
+\tilde{h}_{54}(\phi,X_i)A_\rho \nabla_\sigma \phi\,,
\label{effmet}
\ea
where the effective metrics 
$\mathcal{G}_{\rho\sigma}^{h_5}$ 
and $\mathcal{G}_{\rho\sigma}^{\tilde{h}_5}$ contain 
possible combinations of $g_{\rho\sigma}, A_{\rho}$, 
and $\nabla_{\rho} \phi$. 
The functions $h_{5j}$ and $\tilde{h}_{5j}$ 
(where $j=1,2,3,4$)  depend on 
$\phi$ and $X_1, X_2, X_3$. 
For arbitrary curved backgrounds, the dependence of either 
$X_1$ or $X_3$ in $h_{5j}$ and $\tilde{h}_{5j}$ should 
appear dominantly to ensure that the temporal component of 
$A_{\mu}$ remains non-dynamical \cite{Heisenberg18}. 
On the isotropic and homogenous cosmological background 
this dynamical property of $A_0$ does not manifest 
itself, so we do not restrict the $X_i$ dependence 
in the functions $h_{5j}$ and $\tilde{h}_{5j}$.

The 4-rank tensors $\mathcal{M}_6^{\mu\nu\alpha\beta}$ 
and $\mathcal{N}^{\mu\nu\alpha\beta}_6$ are 
defined, respectively, by 
\be
\mathcal{M}^{\mu\nu\alpha\beta}_6
=2f_{6,X_1} (\phi, X_1) 
\tilde{F}^{\mu\nu}\tilde{F}^{\alpha\beta}\,,
\qquad
\mathcal{N}^{\mu\nu\alpha\beta}_6
=\frac12\tilde{f}_{6,X_3} (\phi, X_3) 
\tilde{F}^{\mu\nu}\tilde{F}^{\alpha\beta}\,.
\ee
The Lagrangian ${\cal L}_{\rm SVT}^{(6)}$ corresponds to 
intrinsic vector modes that vanish in the scalar limit 
$A_{\mu} \to \nabla_{\mu} \pi$. 
In summary, the functional dependence of 
$F, Y_1, Y_2, Y_3$ in $f_2$ and the functions 
$h_{5j}, \tilde{h}_{5j}, f_6, \tilde{f}_6$ accommodate
interactions of intrinsic vector modes.

In the action (\ref{action}), we focused on the interactions 
invariant under the parity transformation ${\cal P}:{\vec x}\to-{\vec x}$. 
In other words, we did not take into account 
the dependence of parity-violating terms like  
$\tilde{F}=-F_{\mu \nu} \tilde{F}^{\mu \nu}/4$ 
in $f_2$. These parity-violating terms 
generate left-handed and right-handed helicity contributions
to the vector perturbation equation, which 
makes the analysis more involved.
The analysis containing parity-violating terms is 
left for a future separate work.

The action of scalar-tensor interactions, which corresponds 
to Horndeski theories \cite{Horndeski,Horn2}, is given by 
\be
{\cal S}_{\rm ST}
=\int d^4x \sqrt{-g} \sum_{n=3}^{5} 
{\cal L}_{\rm ST}^{(n)}\,,
\label{Hoaction}
\ee
with the Lagrangians
\ba
{\cal L}_{\rm ST}^{(3)} &=& 
G_3(\phi,X_1) \square \phi\,,\\
{\cal L}_{\rm ST}^{(4)} &=& 
G_4(\phi,X_1)R+G_{4,X_1}(\phi,X_1) 
\left[ (\square \phi)^2 -(\nabla_{\mu}\nabla_{\nu}\phi) 
(\nabla^{\mu}\nabla^{\nu}\phi) \right]\,,\\
{\cal L}_{\rm ST}^{(5)} 
&=&
G_5(\phi,X_1) G_{\mu \nu} (\nabla^{\mu} \nabla^{\nu} \phi) \nonumber \\
& &
-\frac{1}{6}G_{5,X_1}(\phi,X_1) \left[ (\square \phi)^3 
-3  (\square \phi) (\nabla_{\mu}\nabla_{\nu}\phi) 
(\nabla^{\mu}\nabla^{\nu}\phi)
+2(\nabla^{\mu}\nabla_{\alpha}\phi ) 
(\nabla^{\alpha}\nabla_{\beta}\phi ) 
(\nabla^{\beta}\nabla_{\mu}\phi )\right]\,,
\label{STaction}
\ea
where 
$\square \phi=g^{\mu \nu}
\nabla_{\mu}\nabla_{\nu}\phi$.
The quadratic Lagrangian ${\cal L}_{\rm ST}^{(2)}$ 
of the form $G_2(\phi,X_1)$ is already included 
in the function $f_2$ of the SVT Lagrangian 
${\cal L}_{\rm SVT}^{(2)}$.

For the matter sector, we take into account a perfect fluid 
minimally coupled to gravity. 
This can be described by the Schutz-Sorkin 
action \cite{Sorkin,DGS,GPcosmo}:
\be
{\cal S}_m=-\int d^{4}x \left[ \sqrt{-g}\,\rho_m(n)
+J^{\mu}(\partial_{\mu}\ell+\mathcal{A}_1
\partial_{\mu}\mathcal{B}_1+\mathcal{A}_2
\partial_{\mu}\mathcal{B}_2) \right]\,,
\label{Spf}
\ee
where $J^\mu$ and $\ell$ describe scalar modes, 
while the contributions of vector modes are encoded in 
$\mathcal{A}_{1,2}$ and $\mathcal{B}_{1,2}$. 
The Schutz-Sorkin action has an advantage of appropriately 
dealing with vector perturbations on the FLRW background.
The fluid density $\rho_m$ is a function of 
its number density $n$ defined by 
\be
n=\sqrt{\frac{J^{\mu}J^{\nu}g_{\mu \nu}}{g}}\,.
\label{ndef}
\ee
Varying the action (\ref{Spf}) with respect to $J^{\mu}$, 
it follows that 
\be
u_\mu\equiv\frac{J_\mu}{n\sqrt{-g}}
=\frac1{\rho_{m,n}}\,(\partial_{\mu}\ell+\mathcal{A}_1
\partial_{\mu}\mathcal{B}_1+\mathcal{A}_2\partial_{\mu}\mathcal{B}_2)\,,
\label{umudef}
\ee
where $u_{\mu}$ is the normalized four-velocity, and  
$\rho_{m,n}$ is defined by 
$\rho_{m,n} \equiv \partial\rho_m/\partial n$.

In this paper, we study the cosmology of full SVT theories 
with parity invariance given by the action 
\be
{\cal S}={\cal S}_{\rm SVT}+{\cal S}_{\rm ST}
+{\cal S}_m\,.
\label{actionfull}
\ee
Our analysis can accommodate both Horndeski and generalized 
Proca theories as specific cases. 
The Horndeski theories are characterized by the functions
\be
f_2=f_2(\phi, X_1)\,,\quad f_3=\tilde{f}_3=f_4=f_5
=f_6=\tilde{f}_6
=0\,,\quad h_{5j}=\tilde{h}_{5j}=0\,.
\ee
The generalized Proca theories, which are given by the Lagrangians 
(2.2)-(2.6) of Ref.~\cite{Geff16}, correspond to 
\ba
& &
f_2=f_2(X_3, F, Y_3)\,,\quad 
f_3=f_3(X_3)\,,\quad \tilde{f}_3=0\,,\quad 
f_4=f_4(X_3)\,,\quad f_5=f_5(X_3)\,,\quad 
f_6=0\,,\quad \tilde{f}_6=\tilde{f}_6(X_3)\,,
\nonumber \\
& &
h_{5j}=0\,,\quad \tilde{h}_{51}=-\frac{1}{2} g_5(X_3)\,,
\quad \tilde{h}_{52}=\tilde{h}_{53}=\tilde{h}_{54}
=0\,,\quad 
G_3=G_4=G_5=0\,.
\ea
We note that our SVT theories consist of one scalar 
field $\phi$ and one vector field $A_{\mu}$ coupled to 
gravity, so they do not accommodate scalar bi-galileons and 
their generalizations studied in Refs.~\cite{biGa}.

We will discuss how the background and scalar perturbation equations 
of motion in these two particular theories can be recovered in our general framework.

\section{Background equations of motion}
\label{backsec}

To derive the background equations of motion on the flat FLRW 
background, we take the line element
\be
ds^2=-N^2(t) dt^2+a^2(t) \delta_{ij}dx^i dx^j\,,
\label{FLRW}
\ee
where the lapse $N$ and scale factor $a$ 
depend on the cosmic time $t$.
We consider a time-dependent scalar field 
$\phi(t)$ and a vector field $A_{\mu} (t)$ 
with a nonvanishing temporal component $A_0$ of the form 
$A_{\mu}(t)=\left( A_0(t) N(t), 0, 0, 0 \right)$. 
Then, the quantities defined in Eq.~(\ref{X123}) 
reduce to  
\be
X_1=\frac{\dot{\phi}^2}{2N^2}\,,\qquad 
X_2=\frac{\dot{\phi}A_0}{2N}\,,\qquad 
X_3=\frac{A_0^2}{2}\,,
\ee
where a dot represents a derivative with respect to $t$. 
All the quantities associated with intrinsic vector modes, like 
those defined in Eq.~(\ref{FY123}), 
vanish on the flat FLRW background.

{}From Eq.~(\ref{ndef}), the temporal component $J^{0}$ 
corresponds to the total fluid number ${\cal N}_0$, i.e., 
\be
J^{0}={\cal N}_0=n_0a^3\,,
\ee
where $n_0$ is the background value of $n$. 
On the background (\ref{FLRW}) the vector modes do not 
contribute to the matter action (\ref{Spf}), so that  
\be
\bar{{\cal S}}_m=-\int d^4 x \left( N a^3 \bar{\rho}_m+n_0a^3 \dot{\bar{\ell}} 
\right)\,,
\ee
where a bar is used to represent background values. 

\subsection{Full SVT theories}

On the flat FLRW spacetime (\ref{FLRW}), we compute the action 
(\ref{action}) and vary it with respect to $N$, $a$, 
$\phi$, and $A_0$, and finally set $N=1$. 
Then, the resulting background equations of motion are
\ba
& &
6\left( f_4+G_4 \right)H^2 +f_2 -\dot{\phi}^2 f_{2,X_1}-\frac{1}{2} 
\dot{\phi}A_0 f_{2,X_2}+\dot{\phi}^2 \left( 3H \dot{\phi}G_{3,X_1} 
-G_{3,\phi} \right)+6H \left( \dot{\phi}f_{4,\phi}
-HA_0^2 f_{4,X_3} \right) \nonumber \\
& &
+6 H \dot{\phi} \left( G_{4,\phi}+\dot{\phi}^2 G_{4,X_1 \phi}
-2H \dot{\phi}G_{4,X_1}-H \dot{\phi}^3 G_{4,X_1 X_1} \right)
+2A_0H^2 \left( 3\dot{\phi}f_{5,\phi}-H A_0^2 f_{5,X_3} \right)  \nonumber \\
& &
+H^2 \dot{\phi}^2 \left( 9 G_{5,\phi}+3\dot{\phi}^2 G_{5,X_1 \phi} 
-5H \dot{\phi}G_{5,X_1}-H \dot{\phi}^3 G_{5,X_1 X_1} \right)=\rho_m\,,
\label{back1}\\
& &
2q_t \dot{H}-D_6 \ddot{\phi}+\frac{w_2}{A_0} \dot{A}_0+D_7 \dot{\phi}
=-\rho_m-P_m\,,
 \label{back2}\\
& & 
3D_6 \dot{H}+2D_1 \ddot{\phi}-D_8 \dot{A}_0
+3D_7 H-D_9 A_0-D_5=0\,,
 \label{back3}\\
& &
2 \left( f_{2,X_3}+6H^2f_{4,X_3}-6H \dot{\phi} f_{4,X_3 \phi} 
\right)A_0-2 \left( 6H f_{3,X_3}+6H \tilde{f}_3
+2\dot{\phi}\tilde{f}_{3,\phi}-3H^3 f_{5,X_3}
+3H^2 \dot{\phi}f_{5,X_3 \phi} \right)A_0^2 \nonumber \\
\hspace{-0.3cm}
& &
+12H^2f_{4,X_3X_3} A_0^3 +2H^3 f_{5,X_3X_3} A_0^4
+\left( f_{2,X_2}+4f_{3,\phi}-6H^2 f_{5,\phi} \right)
\dot{\phi}=0\,,
 \label{back4}
\ea
where $H=\dot{a}/a$ is the Hubble expansion rate. 
The matter pressure is given by 
\be
P_m=n_0 \rho_{m,n}-\rho_m\,,
\label{Pm}
\ee
where we used the relation $\dot{\ell}=-\rho_{m,n}$ and 
omitted the bar from the background quantities. 
The quantity $q_t$ in Eq.~(\ref{back2}), which is associated with the no-ghost 
condition of tensor perturbations discussed later, is given by 
\be
q_t=2f_4+2G_4-2A_0^2 f_{4,X_3}-2\dot{\phi}^2 G_{4,X_1}
+A_0 \dot{\phi} f_{5,\phi}-HA_0^3 f_{5,X_3}
+\dot{\phi}^2 G_{5,\phi}-H \dot{\phi}^3 G_{5,X_1}\,.
\label{qt}
\ee
The coefficients $D_1,D_5,D_6,D_7,D_8,D_9$ and $w_2$ 
in Eqs.~(\ref{back2}) and (\ref{back3}) are presented in Appendix~\ref{coeff}. 
As we see later, they also appear as coefficients in the second-order 
action of scalar perturbations. 
The quantity $w_2$ is proportional to $A_0$, so the term $w_2/A_0$  
in Eq.~(\ref{back2}) is not divergent in the limit that $A_0 \to 0$.
We note that Eq.~(\ref{back2}) has been derived by 
varying the action with respect to $a$ and then subtracting 
the corresponding equation of motion from Eq.~(\ref{back1}). 
The matter sector satisfies the continuity equation 
\be
\dot{\rho}_m+3H \left( \rho_m+P_m \right)=0\,,
\label{coneq}
\ee
which are consistent with Eqs.~(\ref{back1})-(\ref{back4}).

Taking the time derivative of Eq.~(\ref{back4}), we obtain
\be
-\frac{3w_2}{A_0} \dot{H}-D_8 \ddot{\phi}
+\frac{2w_5}{A_0^2} \dot{A}_0-D_9 \dot{\phi}=0\,,
\label{back5}
\ee
where $w_5$ is given in Appendix~\ref{coeff}. 
As long as the condition
\be
{\cal D} \equiv 2 \left( 4D_1q_t w_5+3D_1w_2^2
+3D_6^2w_5-A_0^2 D_8^2 q_t-3A_0D_6D_8 w_2
\right) \neq 0
\label{calD}
\ee
is satisfied, the dynamical system is closed.
In other words, we can solve Eqs.~(\ref{back2}), (\ref{back3}), and (\ref{back5}) for $\dot{H}$, 
$\ddot{\phi}$, and $\dot{A}_0$ in the forms:
\ba
\dot{H} 
&=& \frac{1}{{\cal D}} [ A_0^2 D_8 (D_6 D_9 \dot{\phi}
+D_7 D_8 \dot{\phi}-D_9 w_2 )
-A_0 (2D_1 D_9 \dot{\phi}w_2-3D_7D_8H w_2 
+D_5 D_8 w_2-2D_6 D_9 w_5) \nonumber \\
& &~~~-2w_5 (2D_1D_7 \dot{\phi}+3D_6 D_7 H-D_5 D_6)
+(A_0^2 D_8^2-4D_1 w_5)(\rho_m+P_m) ]\,,
\label{dotH} \\
\ddot{\phi}
&=& \frac{1}{{\cal D}} [2A_0^2 D_8 D_9 q_t \dot{\phi} 
+A_0 (3D_6 D_9 \dot{\phi}w_2-3D_7 D_8 \dot{\phi}w_2
+4D_9 q_t w_5+3D_9 w_2^2)
+2w_5 (3D_6D_7 \dot{\phi}-6D_7 H q_t+2D_5 q_t)
\nonumber \\
& &~~~+3w_2^2(D_5-3D_7 H)
-3(A_0 D_8 w_2-2D_6 w_5)(\rho_m+P_m)]\,,
\label{ddotphi} \\
\dot{A}_0 
&=&  \frac{A_0}{{\cal D}} [A_0 D_9 (2A_0 D_8 q_t
+4D_1 q_t \dot{\phi}+3D_6^2 \dot{\phi})
+A_0 (3D_6 D_7 D_8 \dot{\phi}-6D_7 D_8 H q_t 
+2D_5 D_8 q_t+3D_6 D_9 w_2)\nonumber \\
& &~~~-3w_2 (2D_1 D_7 \dot{\phi}+3D_6 D_7 H-D_5 D_6)
+3 (A_0 D_6 D_8-2D_1 w_2)(\rho_m+P_m)]\,.
\label{dotA0}
\ea
The initial conditions of $H, \dot{\phi}, A_0$ 
should be chosen to be consistent with 
Eqs.~(\ref{back1}) and (\ref{back4}).
In Sec.~\ref{scasec}, we show that the determinant (\ref{calD}) 
is related to a quantity $q_s$ associated with 
the no-ghost condition of scalar perturbations. 

We introduce the dark energy density $\rho_{\rm DE}$ and 
pressure $P_{\rm DE}$ in the forms:
\ba
\rho_{\rm DE}
&=& \frac{3H^2}{8\pi G}-\rho_m\,,
\label{rhode}\\
P_{\rm DE}
&=& -\frac{2\dot{H}+3H^2}{8\pi G}-P_m\,,
\label{Pde}
\ea
where $G$ is the Newton gravitational constant. 
Then, the dark sector obeys the continuity equation 
\be
\dot{\rho}_{\rm DE}+3H \left( \rho_{\rm DE}
+P_{\rm DE} \right)=0\,,
\ee
where we used Eq.~(\ref{coneq}).
We can explicitly compute $\rho_{\rm DE}$ and $P_{\rm DE}$
by solving Eqs.~(\ref{back1})-(\ref{back2}) for 
$\rho_m, P_m$ and substitute them into 
Eqs.~(\ref{rhode})-(\ref{Pde}). 
To calculate $\dot{H}, \ddot{\phi}, \dot{A}_0$ appearing 
in the expression of $P_{\rm DE}$ explicitly, 
we need to employ Eqs.~(\ref{dotH})-(\ref{dotA0}).

\subsection{Horndeski theories}

In Horndeski theories the temporal vector component $A_0$ 
is absent, so Eqs.~(\ref{back4}) and (\ref{back5}) are redundant.
In this case, as long as the condition 
\be
{\cal D}_{\rm Ho} \equiv 4D_1 q_t+3D_6^2 \neq 0
\label{Dho}
\ee
is satisfied, we can solve Eqs.~(\ref{back2}) and 
(\ref{back3}) for $\dot{H}$ and $\ddot{\phi}$, as 
\ba
\dot{H} 
&=& -\frac{1}{{\cal D}_{\rm Ho}} 
\left[ 2D_1 D_7 \dot{\phi}-D_6 (D_5-3D_7 H)
+2D_1 (\rho_m+P_m) \right]\,,\label{Ho1} \\
\ddot{\phi}
&=& \frac{1}{{\cal D}_{\rm Ho}} \left[ 
3D_6 D_7 \dot{\phi}+2 (D_5-3D_7H)q_t
+3D_6 (\rho_m+P_m) \right]\,.\label{Ho2}
\ea
The determinant ${\cal D}_{\rm Ho}$ is proportional to a quantity $q_{s,{\rm Ho}}$ associated with the no-ghost condition of scalar perturbations discussed later 
in Sec.~\ref{scasec}.

\subsection{Generalized Proca theories}

In generalized Proca theories the scalar field $\phi$ is absent,
in which case Eq.~(\ref{back3}) is redundant. 
Provided that the condition 
\be
{\cal D}_{\rm GP} \equiv 4w_5 q_t+3w_2^2 
\neq 0
\label{DGP}
\ee
is satisfied, we can solve Eqs.~(\ref{back2}) and 
(\ref{back5}) for $\dot{H}$ and $\dot{A}_0$, as 
\ba
\dot{H} 
&=& -\frac{2w_5}{{\cal D}_{\rm GP}} 
\left( \rho_m+P_m \right)\,,
\label{GP1}\\
\dot{A}_0
&=&
-\frac{3w_2}{{\cal D}_{\rm GP}}A_0
\left( \rho_m+P_m \right)\,.
\label{GP2}
\ea
The determinant ${\cal D}_{\rm GP}$ is related to a quantity 
$q_{s,{\rm GP}}$ associated with the no-ghost condition 
of scalar perturbations.

\section{Tensor and vector perturbations}
\label{TVsec}

We proceed to the study of linear cosmological perturbations in 
SVT theories given by the action (\ref{actionfull}). 
In doing so, we decompose the perturbations into tensor, vector, 
and scalar modes on the flat FLRW background \cite{Bardeen}.
We consider the perturbed line element  in the flat gauge:
\be
ds^2=-(1+2\alpha) dt^2+2 
\left(\partial_i \chi+V_i \right) dt dx^i 
+a^2(t) \left( \delta_{ij}+h_{ij} \right) dx^idx^j\,,
\label{permet}
\ee
where $\alpha$ and $\chi$ are scalar perturbations, with 
the notation $\partial_i \chi \equiv \partial \chi/\partial x^i$.
The vector perturbation $V_i$ obeys the transverse condition
\be
\partial^i V_i=0\,,
\label{vectra}
\ee
whereas the tensor perturbation $h_{ij}$ satisfies the 
transverse and traceless conditions 
\be
\partial^j h_{ij}=0\,,\qquad {h_i}^i=0\,.
\label{tentra}
\ee

For the scalar field $\phi$ and the vector field $A^{\mu}$, 
we decompose them into the background and perturbed 
parts, as
\ba
\phi&=&\bar{\phi}(t)+\delta \phi\,,\\
A^{0}&=&-\bar{A}_0(t)+\delta A\,,\qquad 
A_i=\partial_i \psi+Z_i\,,
\ea
where $\delta \phi, \delta A, \psi$ are scalar perturbations, 
and $Z_i$ is the vector perturbation obeying
\be
\partial^i Z_i=0\,.
\label{vectra2}
\ee
In the following, we omit the bar from the background 
quantities.

For the matter sector, the temporal and spatial components 
of $J^{\mu}$ in Eq.~(\ref{Spf}) are decomposed as
\ba
J^{0} =  \mathcal{N}_{0}+\delta J\,,\qquad
J^{k} =\frac{1}{a^2(t)}\,\delta^{ki}
\left( \partial_{i}\delta j+W_i \right)\,,
\label{elldef}
\ea
where $\delta J$ and $\delta j$ correspond to 
scalar perturbations, and $W_i$ is the vector 
perturbation satisfying
\be
\partial^i W_i=0\,.
\label{vectra3}
\ee
Without loss of generality, we can choose the vector fields 
$V_i, Z_i, W_i$ obeying the transverse conditions 
(\ref{vectra}), (\ref{vectra2}), and (\ref{vectra3}) 
in the forms
\be
V_i=\left( V_1(t,z), V_2(t,z), 0 \right)\,,\qquad
Z_i=\left( Z_1(t,z), Z_2(t,z), 0 \right)\,,\qquad
W_i=\left( W_1(t,z),W_2(t,z),0 \right)\,,
\label{Wi}
\ee
whose nonvanishing components depend on $t$ 
and the third spatial coordinate $z$. 

The scalar quantity $\ell$ is expressed as  
\be
\ell=-\int^{t} \rho_{m,n} 
(\tilde{t})d\tilde{t} 
-\rho_{m,n}v\,,
\label{ells}
\ee
where $v$ is the velocity potential. 
The quantities ${\cal A}_1,{\cal A}_2,{\cal B}_1,{\cal B}_2$, 
which are related to intrinsic vector modes, 
can be chosen as \cite{DGS,GPcosmo}
\be
{\cal A}_1=\delta {\cal A}_1(t,z)\,,\qquad  
{\cal A}_2=\delta {\cal A}_2(t,z)\,,\qquad 
{\cal B}_1=x+\delta {\cal B}_1(t,z)\,,\qquad
{\cal B}_2=y+\delta {\cal B}_2(t,z)\,,\label{ABi}
\ee
where $\delta {\cal A}_{1,2}$ and $\delta {\cal B}_{1,2}$ 
are perturbed quantities that depend on $t$ and $z$. 
Substituting Eq.~(\ref{ells}) into (\ref{umudef}), the spatial 
component of $u_{\mu}$ is expressed in the form 
\be
u_i=-\partial_i v+v_i\,,
\ee
where the vector components $v_i$ (with $i=1,2$) 
are related to $\delta {\cal A}_i$, as
\be
\delta {\cal A}_i=\rho_{m,n}v_i\,.
\label{AiVi}
\ee
The transverse condition $\partial^iv_i=0$ is satisfied for 
${\cal A}_i$ given in Eq.~(\ref{ABi}). 

\subsection{Tensor perturbations}

We first compute the second-order action of tensor perturbations $h_{ij}$. 
To satisfy the transverse and traceless conditions (\ref{tentra}), 
we choose nonvanishing components of $h_{ij}$ in the forms
\be
h_{11}=h_1(t,z)\,,\qquad 
h_{22}=-h_1(t,z)\,,\qquad
h_{12}=h_{21}= h_2(t,z)\,,
\ee
where the functions $h_1$ and $h_2$ characterize
two polarization states of the tensor sector.

Expanding the action ${\cal S}_{\rm SVT}
+{\cal S}_{\rm ST}$ up to second order in 
perturbations and integrating it by parts, the 
quadratic action contains the terms 
$\dot{h}_i^2$, $(\partial h_i)^2$, $h_i^2$ 
(where $i=1,2$).  The second-order action 
of  ${\cal S}_m$ associated with the tensor sector
can be written in the form 
$({\cal S}_m^{(2)})_t=-\int dtd^3x [(\sqrt{-g})^{(2)} \rho_m
+\sqrt{-\bar{g}} \rho_{m,n}\delta n]$, where 
$(\sqrt{-g})^{(2)}=-a^3(h_1^2+h_2^2)/2$ 
and $\delta n=n_0 (h_1^2+h_2^2)/2$ 
with $\sqrt{-\bar{g}}=a^3$. 
Then, it follows that 
\be
({\cal S}_m^{(2)})_t=-\int dt d^3 x \sum_{i=1}^{2} 
\frac{1}{2}a^3 P_m h_i^2\,,
\ee
where $P_m$ is the matter pressure given by Eq.~(\ref{Pm}).
Taking into account the contribution ${\cal S}_m^{(2)}$ 
to the second-order action of ${\cal S}_{\rm SVT}
+{\cal S}_{\rm ST}$ and eliminating $P_m$ by using the background Eqs.~(\ref{back1}) and (\ref{back2}), 
the terms proportional to $h_i^2$ identically vanish. 
Then, the second-order action of 
tensor perturbations yields 
\be
{\cal S}_t^{(2)}=\int dt d^3x \sum_{i=1}^{2}
\frac{a^3}{4}q_t \left[ \dot{h}_i^2-\frac{c_t^2}{a^2} 
(\partial h_i)^2 \right]\,,
\ee
where $q_t$ is given by Eq.~(\ref{qt}), and
\be
c_t^2=\frac{2f_4+2G_4-A_0\dot{\phi}f_{5,\phi}
-\dot{A}_0 A_0^2 f_{5,X_3}-\dot{\phi}^2G_{5,\phi} 
-\dot{\phi}^2 \ddot{\phi} G_{5,X_1}}
{2f_4+2G_4-2A_0^2 f_{4,X_3}-2\dot{\phi}^2 G_{4,X_1}
+A_0 \dot{\phi} f_{5,\phi}-HA_0^3 f_{5,X_3}
+\dot{\phi}^2 G_{5,\phi}-H \dot{\phi}^3 G_{5,X_1}}\,.
\label{ct}
\ee
The quantity $q_t$ is associated with the no-ghost condition 
of tensor perturbations, while $c_t^2$ is the propagation 
speed squared of gravitational waves relevant to the Laplacian instability.
To avoid the ghost and Laplacian instabilities, 
we require that 
\be
q_t>0\,,\qquad c_t^2> 0\,.
\label{qtcon}
\ee
Taking the limits $f_4, f_{4,X_3}, f_{5,\phi}, f_{5,X_3} \to 0$  
in Eqs.~(\ref{qt}) and (\ref{ct}), we recover the values of 
$q_t$ and $c_t^2$ in Horndeski theories \cite{Horn2,exga}. 
The values of $q_t$ and $c_t^2$ in generalized 
Proca theories \cite{GPcosmo}
also follow by taking the limits 
$G_4, G_{4,X_1}, G_{5,\phi}, G_{5,X_1}, f_{5,\phi} \to 0$.

Applying the SVT theories to today's universe, there is a tight 
bound on $c_t$ constrained from the GW170817 
event \cite{GW170817} together with the 
electromagnetic counterpart \cite{Goldstein}: 
\be
-3\times10^{-15}\leq c_t-1 \leq 7\times10^{-16}\,.
\ee
If we strictly demand $c_t^2=1$ in Eq.~(\ref{ct}), 
the SVT theories need to satisfy the condition 
\be
2A_0^2 f_{4,X_3}-2A_0\dot{\phi}f_{5,\phi}
+A_0^2 \left( HA_0 
-\dot{A}_0 \right)f_{5,X_3}
+2\dot{\phi}^2 G_{4,X_1}
-2\dot{\phi}^2 G_{5,\phi}
+\dot{\phi}^2 \left( H \dot{\phi} -\ddot{\phi} 
\right) G_{5,X_1}=0\,.
\label{ctre}
\ee
If we consider the case in which each term on the left hand 
side of Eq.~(\ref{ctre}) exactly vanishes without the 
cancellation between different terms, the couplings are 
constrained to be 
\be
f_4=f_4(\phi)\,,\qquad f_5={\rm constant}\,,\qquad 
G_4=G_4(\phi)\,,\qquad G_5={\rm constant}\,.
\label{ctcon}
\ee
The Lagrangians up to cubic order as well as intrinsic 
vector modes like ${\cal L}_{\rm SVT}^{(6)}$
do not affect the tensor propagation speed.

We can also consider cases in which some of nonvanishing terms on the left hand side of Eq.~(\ref{ctre}) cancel each other. In scalar-tensor theories 
beyond Horndeski gravity, it is also possible to construct 
similar tuned cosmological models with $c_t^2=1$ [39,41,43,44]. 
In our case, one of the simplest examples consistent with 
Eq.~(4.20) is the functions
$f_{4,X_3}=1/X_3$ and $G_{4,X_1}=-1/X_1$ with constant 
couplings $f_5$ and $G_5$.
Of course, the fact that $c_t^2=1$ alone does not guarantee 
the stability of theories against vector and scalar perturbations, 
so we need to confirm whether such theories 
satisfy all the stability criteria required for the 
cosmological viability. 
In Sec.~\ref{conmodelsec}, we will consider SVT theories 
given by the couplings (\ref{ctcon}), leaving the analysis 
of more general cases with $c_t^2=1$ as a future work.

\subsection{Vector perturbations}

Let us proceed to the derivation of second-order 
action of vector perturbations.
Expanding the matter action ${\cal S}_m$ in terms of intrinsic 
vector perturbations $W_i, \delta {\cal A}_i, \delta {\cal B}_i$, 
and $V_i$, the resulting second-order action 
is given by \cite{GPcosmo,Geff16}
\be
({\cal S}_{m}^{(2)})_v
= \int dtd^3x  \sum_{i=1}^{2}
\biggl[ \frac{1}{2a^2{\cal N}_0} \left\{
\rho_{m,n} \left (W_i^2+{\cal N}_0^2\, V_i^2 \right)
+{\cal N}_0 \left(2\rho_{m,n}V_iW_i
-a^3\rho_m V_i^2 \right)\right\} 
-{\cal N}_0 \delta {\cal A}_i \dot{\delta \cal B}_i
-\frac{1}{a^2} W_i \delta {\cal A}_i \biggr]\,.
\label{Lvsc}
\ee
Since the second-order action of 
${\cal S}_{\rm SVT}+{\cal S}_{\rm ST}$ does not 
contain the perturbations 
$W_i, \delta {\cal A}_i, \delta {\cal B}_i$, 
the action (\ref{Lvsc}) can be independently varied with 
respect to $W_i, \delta {\cal A}_i, \delta {\cal B}_i$.
This leads to the following relations 
\ba
W_i &=&{\cal N}_0\,(v_i-V_i)\,,\\
v_i &=& V_i-a^2\dot{\delta {\cal B}}_i\,,
\label{vi}\\
\delta {\cal A}_i &=&\rho_{m,n}v_i=C_i\,,
\label{calA}
\ea
where $C_i$ ($i=1,2$) are constants in time.
After integrating out the perturbations $W_i$ and $\delta A_i$, 
the matter action (\ref{Lvsc}) reduces to
\be
({\cal S}_{m}^{(2)})_v=\int dtd^3x \sum_{i=1}^{2}
\frac{a}{2} \left[ \left( \rho_m+P_m \right)
v_i^2-\rho_m V_i^2 \right]\,,
\label{LVM2}
\ee
where $v_i$ contains $V_i$ through Eq.~(\ref{vi}).

Taking into account the contribution (\ref{LVM2}) to 
the second-order action of ${\cal S}_{\rm SVT}
+{\cal S}_{\rm ST}$ and using the background equations 
of motion, the quadratic action 
of vector perturbations reads
\be
{\cal S}_v^{(2)}=\int dt d^3 x \sum_{i=1}^2 
\left[ \frac{aq_v}{2} \dot{Z}_i^2-\frac{1}{2a} 
\alpha_1 (\partial Z_i)^2-\frac{a}{2} \alpha_2 Z_i^2
+\frac{1}{2a} \alpha_3 (\partial V_i)(\partial Z_i) 
+\frac{q_t}{4a} (\partial V_i)^2
+\frac{a}{2}  \left( \rho_m +P_m \right)v_i^2 \right]\,,
\label{Sv}
\ee
where 
\ba
q_v &=& f_{2,F}+2\dot{\phi}^2 f_{2,Y_1}
+2\dot{\phi}A_0 f_{2,Y_2}+2A_0^2f_{2,Y_3}
-4H \left( \dot{\phi} h_{51}+2A_0  \tilde{h}_{51} \right)
+8H^2 \left( f_6+\tilde{f}_6+\dot{\phi}^2 f_{6,X_1}
+A_0^2 \tilde{f}_{6,X_3} \right)\,,\label{qv}\\
\alpha_1 &=& f_{2,F} - 4\dot{A}_0\tilde{h}_{51} 
+8\left( H^2+\dot{H} \right) \left( f_6+\tilde{f}_6 \right)
-2\ddot{\phi}\,h_{51}
+H \biggl[ 2\dot{\phi} \left( \dot{\phi}^2 h_{52}
-h_{51}+4\ddot{\phi}f_{6,X_1} \right)  \nonumber \\
& &
-A_0 \left\{
4\tilde{h}_{51}-2\dot{\phi}^2
\left( h_{54}+2\tilde{h}_{52} \right) 
-8\dot{A}_0 \tilde{f}_{6,X_3}\right\}
+2\dot{\phi}A_0^2(h_{53}+2\tilde{h}_{54})
+4A_0^3 \tilde{h}_{53}
\biggr]\,,\\
\alpha_2 &=&
f_{2,X_3}+4\dot{H}f_{4,X_3}
-2 \left( \dot{A}_0+3HA_0 \right) 
\left( f_{3,X_3}+\tilde{f}_3 \right)
-2\dot{\phi}A_0 \tilde{f}_{3,\phi}
+2H ( 3H f_{4,X_3}+
3HA_0^2 f_{4,X_3 X_3}+2A_0 \dot{A}_0
f_{4,X_3X_3} \nonumber \\
& &-\dot{\phi}f_{4,X_3 \phi})
+H \left( H \dot{A}_0+2 \dot{H} A_0+3H^2 A_0
\right) f_{5,X_3}+H^2 A_0 \left( 
H A_0^2 f_{5,X_3 X_3}+A_0 \dot{A}_0 
f_{5,X_3 X_3}-2\dot{\phi}f_{5,X_3 \phi}\right)\,,\\
\alpha_3 &=& 
-2A_0 f_{4,X_3}-HA_0^2 f_{5,X_3}+\dot{\phi}f_{5,\phi}\,.
\ea
Apart from the last term of Eq.~(\ref{Sv}), all the other terms 
in ${\cal S}_v^{(2)}$ are exactly the same as those derived 
for the theories with the action ${\cal S}_{\rm SVT}$ 
alone \cite{HKT18}. 
This reflects the fact that the action ${\cal S}_{\rm ST}$ 
of scalar-tensor theories does not give rise to any 
modification to the vector sector.

Varying the action (\ref{Sv}) with respect to $V_i$, 
we obtain
\be
\partial^2 \left( \alpha_3 Z_i+q_t V_i \right)
=2a^2 \left( \rho_m+P_m \right)v_i\,.
\label{ViZi}
\ee
On using Eq.~(\ref{calA}), the term on the right hand side 
of Eq.~(\ref{ViZi}) can be expressed as $2a^2 n_0 C_i$. 
Taking the small-scale limit\footnote{Here and in the following, 
we use the word ``small-scale limit'' for the meaning of taking the 
large comoving wavelength limit ($k \to \infty$) in the perturbation 
equations of motion. In a strict sense, this limit can be applied to 
small-scale perturbations in the {\it linear} regime of gravity.} 
in Eq.~(\ref{ViZi}) under 
the condition that  $C_i$ do not  depend on scales, 
it follows that $V_i \simeq -\alpha_3 Z_i/q_t$. 
Since the last term of Eq.~(\ref{Sv}) is irrelevant to 
the dynamics of vector perturbations in the small-scale limit, 
the action (\ref{Sv}) reduces to 
\be
{\cal S}_v^{(2)} \simeq \int dt d^3 x \sum_{i=1}^2 
\frac{a}{2}q_v \left[ \dot{Z}_i^2-\frac{c_v^2}{a^2} 
\left( \partial Z_i \right)^2-\frac{\alpha_2}{q_v}Z_i^2 
\right]\,,
\label{Sv2}
\ee
where 
\be
c_v^2=\frac{2\alpha_1 q_t+\alpha_3^2}{2q_t q_v}\,.
\label{cv}
\ee
Hence there are two dynamically propagating fields $Z_1$ 
and $Z_2$ in the vector sector. 
The ghost and Laplacian instabilities are absent 
under the conditions 
\be
q_v>0\,,\qquad c_v^2>0\,,
\ee
which are exactly the same as those derived for the 
action ${\cal S}_{\rm SVT}$ alone \cite{HKT18}.
This means that neither the action ${\cal S}_{\rm ST}$ 
of scalar-tensor theories nor the matter action ${\cal S}_m$ 
changes the stability conditions of vector perturbations 
in the small-scale limit.

\section{Scalar perturbations}
\label{scasec}

For the scalar sector, there are metric perturbations 
$\alpha, \chi$, scalar-field perturbation $\delta \phi$, 
and perturbations $\delta A, \psi$ arising from the 
vector field. The matter perfect fluid also contains the 
scalar perturbations $\delta J, \delta j, v$. 
We introduce the matter density perturbation 
$\delta \rho_m$, as 
\be
\delta \rho_m=\frac{\rho_{m,n}}{a^3} \delta J\,.
\ee
Defining $\delta \rho_m$ in this way, the perturbation 
of the fluid number density $n$, expanded up to 
second order, yields
\be
\delta n= \frac{\delta \rho_m}{\rho_{m,n}}
-\frac{{\cal N}_0^2 (\partial \chi)^2+2{\cal N}_0
\partial \chi \partial \delta j+(\partial \delta j)^2}
{2{\cal N}_0a^5}\,,
\ee
so that $\delta n$ is equivalent to 
$\delta \rho_m/\rho_{m,n}$ at first order. 
Expanding the Schutz-Sorkin action (\ref{Spf}) up to 
quadratic order in perturbations, we obtain the 
second-order action:
\ba
({\cal S}_m^{(2)})_s
&=&\int dt d^3 x \biggl\{ \frac{1}{2a^5n_0 \rho_{m,n}^2}
\left[ \rho_{m,n} \left( \rho_{m,n}^2 (\partial \delta j)^2
+2a^3n_0 \rho_{m,n}^2 \partial \delta j \partial v
+2a^8n_0  \rho_{m,n}  \dot{v}\,\delta \rho_m
-6a^8n_0^2 \rho_{m,nn}Hv \delta \rho_m \right) 
\right.
\nonumber \\
& &\left. 
\qquad \qquad -a^8 n_0 \rho_{m,nn} (\delta \rho_m)^2 
\right]
-a^3 \alpha \delta \rho_m
+\frac{\rho_{m,n}}{a^2} \partial \chi \partial \delta j
+\frac{1}{2}a^3 \rho_m \alpha^2+\frac{1}{2}
a \left( n_0 \rho_{m,n}-\rho_m \right) (\partial \chi)^2
\biggr\}\,.
\label{SMS}
\ea
Varying this action with respect to $\delta j$, it follows that 
\be
\partial \delta j=-a^3 n_0 \left( \partial v+\partial \chi 
\right)\,.
\ee
On using this relation, we can eliminate the perturbation 
$\delta j$ from Eq.~(\ref{SMS}) to give 
\be
({\cal S}_m^{(2)})_s
=\int dt d^3x\,a^3 \left\{ \left( \dot{v}-3H c_m^2v-\alpha 
\right) \delta \rho_m-\frac{c_m^2 (\delta \rho_m)^2}
{2n_0 \rho_{m,n}}-\frac{n_0 \rho_{m,n}}{2a^2} 
\left[ (\partial v)^2+2\partial v\partial \chi \right]
+\frac{1}{2} \rho_m \alpha^2-\frac{\rho_m}{2a^2} 
(\partial \chi)^2 \right\}\,,
\label{SMS2}
\ee
where $c_m^2$ is the matter sound speed squared defined by 
\be
c_m^2=\frac{P_{m,n}}{\rho_{m,n}}
=\frac{n_0 \rho_{m,nn}}{\rho_{m,n}}\,.
\ee

We also expand the action ${\cal S}_{\rm SVT}
+{\cal S}_{\rm ST}$ up to quadratic order in scalar perturbations and take the sum with 
$({\cal S}_m^{(2)})_s$. Using the background 
Eq.~(\ref{back1}), the last term of Eq.~(\ref{SMS2}) 
cancels out and the term $\rho_m \alpha^2/2$ can be 
absorbed into one of contributions arising from 
${\cal S}_{\rm SVT}+{\cal S}_{\rm ST}$.  
After the integration by parts, the resulting second-order 
action is given by 
\be
{\cal S}_s^{(2)}=\int dt d^3x 
\left({\cal L}_{s1}+{\cal L}_{s2}+{\cal L}_{s3} \right)\,, 
\label{Ss}
\ee
where 
\ba
\hspace{-0.7cm}
{\cal L}_{s1}
&=& a^3\left[
D_1\dot{\dphi}^2+D_2\frac{(\partial\dphi)^2}{a^2}+D_3\dphi^2
+\left(D_4\dot{\dphi}+D_5\dphi+D_6\frac{\partial^2\dphi}{a^2}\right) \alpha
-\left(D_6\dot{\dphi}-D_7\dphi\right)\frac{\partial^2\chi}{a^2}
\right.\notag\\
\hspace{-0.7cm}
&&
\left.~~~
+\left(D_8\dot{\dphi}+D_9\dphi\right)\da+D_{10}\,\dphi\,\frac{\partial^2\psi}{a^2}
\right]\,, \\
\hspace{-0.7cm}
{\cal L}_{s2}
&=& a^3\left[
\left(w_1\alpha-w_2\frac{\da}{A_0}\right)\frac{\partial^2\chi}{a^2}
-w_3\frac{(\partial\alpha)^2}{a^2}+w_4\alpha^2
-\left(w_3\frac{\partial^2\da}{a^2A_0}-w_8\frac{\da}{A_0}
+w_3\frac{\partial^2\dot{\psi}}{a^2A_0}+w_6\frac{\partial^2\psi}{a^2}\right)\alpha
\right.\notag\\
\hspace{-0.7cm}
&&\left.~~~
-w_3\frac{(\partial\da)^2}{4a^2A_0^2}+w_5\frac{\da^2}{A_0^2}
+\left\{w_3\dot{\psi}-(w_2-A_0w_6)\psi\right\}\frac{\partial^2\da}{2a^2A_0^2}
-w_3\frac{(\partial\dot{\psi})^2}{4a^2A_0^2}+w_7\frac{(\partial\psi)^2}{2a^2}
\right]\,,\label{LGP}\\
\hspace{-0.7cm}
{\cal L}_{s3}
&=&a^3 \left[ \left( \rho_m+P_m \right)v\frac{\partial^2 \chi}
{a^2}-v\dot{\delta \rho}_m-3H (1+c_m^2) v\delta \rho_m 
-\frac{1}{2} (\rho_m+P_m) \frac{(\partial v)^2}{a^2}
-\frac{c_m^2}{2 (\rho_m+P_m)} (\delta \rho_m)^2 
-\alpha \delta \rho_m \right].
\ea
In Appendix~\ref{coeff}, we show explicit expressions of the 
coefficients $D_{1,\cdots,10}$ and $w_{1,\cdots,8}$. 
The Lagrangian ${\cal L}_{s1}$ arises from the field perturbation $\delta \phi$, so it vanishes in generalized 
Proca theories. The Lagrangians 
${\cal L}_{s1}$ and ${\cal L}_{s2}$ have the same 
structures as those of SVT theories with
the action ${\cal S}_{\rm SVT}$ alone \cite{HKT18}. 
The difference arises only through the 
coefficients $D_{1,\cdots,10}$ and $w_{1,\cdots,8}$. 
The Lagrangian ${\cal L}_{s3}$ newly 
arises from the matter sector.
The intrinsic vector modes affect the scalar sector 
only through the quantity $w_3=-2A_0^2q_v$.

The second-order action (\ref{Ss}) contains scalar 
perturbations $\alpha, \chi, \delta A, v$ and 
$\psi, \delta \phi, \delta \rho_m$, among which the 
last three quantities correspond to dynamical 
propagating DOFs.
Varying the action (\ref{Ss}) with respect to 
$\alpha, \chi, \delta A, v$, we obtain their 
equations of motion in Fourier space, as
\ba
&&
D_4\dot{\dphi}+D_5\dphi+2w_4\alpha+w_8\frac{\da}{A_0}
+\frac{k^2}{a^2}\left( w_6\psi-w_1\chi-D_6\dphi
-{\cal Y} \right)-\delta \rho_m=0\,,
\label{eqalpha}\\
&&
D_6\dot{\dphi}-D_7\dphi-w_1\alpha
-\left( \rho_m+P_m \right)v
+w_2\frac{\da}{A_0}=0\,,
\label{eqchi}\\
&&
D_8\dot{\dphi}+D_9\dphi+w_8\frac{\alpha}{A_0}+2w_5\frac{\da}{A_0^2}
+\frac{k^2}{a^2}\frac{1}{A_0} 
\left( w_2\chi-\frac{A_0w_6-w_2}{2A_0}\psi
+\frac{1}{2}{\cal Y}
\right)=0\,,
\label{eqdA}\\
& &
\dot{\delta \rho}_m+3 \left( 1+c_m^2 \right)H \delta \rho_m
+\frac{k^2}{a^2} \left( \rho_m+P_m \right) 
\left( v+\chi \right)=0\,,
\label{eqdrho}
\ea
where $k$ is a comoving wavenumber, and 
\be
{\cal Y} \equiv -\frac{w_3}{A_0} 
\left( \dot{\psi}+\delta A-2\alpha A_0 \right)\,.
\label{Ydef}
\ee
Variations of the action (\ref{Ss}) with respect to $\psi, \delta \phi, \delta \rho_m$ 
lead to the following equations of motion:
\ba
& &
\dot{\cal Y}+\left( H -\frac{\dot{A}_0}{A_0} \right){\cal Y}
-\frac{1}{A_0} \left[ (2w_6 \alpha+2w_7 \psi-2D_{10} \delta \phi) 
A_0^2+(w_2-w_6 A_0 )\delta A \right]=0\,,\label{calYeq}\\
& &
\dot{\cal Z}+3H {\cal Z}-2D_3 \delta \phi-D_5 \alpha-D_9 \delta A
-\frac{k^2}{a^2} \left( 2D_2 \delta \phi -D_6 \alpha 
-D_7 \chi-D_{10}\psi \right)=0\,,\label{calZeq}\\
& &
\dot{v}-3Hc_m^2 v-\frac{c_m^2}{\rho_m+P_m} \delta \rho_m 
-\alpha=0\,,
\label{veq}
\ea
where 
\be
{\cal Z} \equiv 2D_1 \dot{\delta \phi}+D_4 \alpha
+D_6 \frac{k^2}{a^2}\chi+D_8 \delta A\,.
\label{Zdef}
\ee
\subsection{Full SVT theories}

In SVT theories, there are three dynamical DOFs 
characterized by the matrix 
\be
\vec{\mathcal{X}}^{t}=\left(\psi, \dphi, 
\frac{\delta \rho_m}{k} \right) \,.
\ee
To eliminate the non-dynamical DOFs from 
the action (\ref{Ss}), we solve Eqs.~(\ref{eqalpha})-(\ref{eqdrho}) for $\alpha, \chi, \delta A, v$ 
and substitute them into Eq.~(\ref{Ss}).
Then, the second-order scalar action can be expressed 
in the form 
\be
{\cal S}_s^{(2)}=\int dt d^3x\,a^{3}\left( 
\dot{\vec{\mathcal{X}}}^{t}{\bm K}\dot{\vec{\mathcal{X}}}
-\frac{k^2}{a^2}\vec{\mathcal{X}}^{t}{\bm G}\vec{\mathcal{X}}
-\vec{\mathcal{X}}^{t}{\bm M}\vec{\mathcal{X}}
-\vec{\mathcal{X}}^{t}{\bm B}\dot{\vec{\mathcal{X}}}
\right)\,,
\label{Ss2}
\ee
where ${\bm K}$, ${\bm G}$, ${\bm M}$, ${\bm B}$ 
are $3 \times 3$ matrices. 
We perform the integrations by parts such that the matrix 
components of neither ${\bm B}$ nor ${\bm M}$
contain the $k^2$ terms for $k \to \infty$. 
Then, in the small-scale limit, the nonvanishing matrix 
components of ${\bm K}$ and ${\bm G}$ are given by 
\ba
&&
K_{11}=\frac{w_1^2w_5+w_2^2w_4+w_1w_2w_8}{A_0^2(w_1-2w_2)^2}\,,\notag\\
&&
K_{22}=D_1+\frac{D_6}{w_1-2w_2}\left(D_4
+\frac{w_4+4w_5+2w_8}{w_1-2w_2}D_6+2A_0D_8\right)
\,,\notag\\
&&
K_{12}=K_{21}=-\frac{1}{2A_0(w_1-2w_2)}\left[w_2D_4
+\frac{w_1(4w_5+w_8)+2w_2(w_4+w_8)}{w_1-2w_2}D_6
+A_0w_1D_8\right]\,,\notag\\
&&
K_{33}=\frac{a^2}{2(\rho_m+P_m)}\,,
\label{Kmat}
\ea
and 
\ba
&&
G_{11}=\dot{E_1}+H E_1-\frac{4A_0^2}{w_3}E_1^2
-\frac{w_7}{2}-\frac{w_2^2(\rho_m+P_m)}
{2A_0^2 (w_1-2w_2)^2}\,,\notag\\
&&
G_{22}=\dot{E_2}+H E_2-\frac{D_6D_7}{w_1-2w_2}
-\frac{4A_0^2}{w_3}E_3^2-D_2
-\frac{D_6^2 (\rho_m+P_m)}{2 (w_1-2w_2)^2}\,,\notag\\
&&
G_{12}=G_{21}=\dot{E_3}+H E_3-\frac{4A_0^2}{w_3}E_1E_3
+\frac{w_2}{2A_0(w_1-2w_2)}D_7+\frac{D_{10}}{2}
+\frac{w_2 D_6(\rho_m+P_m)}{2A_0(w_1-2w_2)^2}\,,
\notag\\
&&
G_{33}=\frac{c_m^2 a^2}{2(\rho_m+P_m)}\,,
\label{Gmat}
\ea
where 
\be
E_1=\frac{w_6}{4A_0}-\frac{w_1w_2}{4A_0^2(w_1-2w_2)}\,,\qquad 
E_2=-\frac{D_6^2}{2(w_1-2w_2)}\,,\qquad 
E_3=\frac{w_2D_6}{2A_0(w_1-2w_2)}\,.
\label{E123}
\ee
These expressions are valid for the SVT theories with 
$A_0 \neq 0$, $w_1-2w_2 \neq 0$, and $w_3 \neq 0$.
Since the off-diagonal components $K_{13}, K_{23}, G_{13}, G_{23}$ 
vanish, the matter perturbation $\delta \rho_m$ is decoupled from 
other fields $\psi$ and $\delta \phi$. 
Provided that the two conditions 
\be
\rho_m+P_m>0\,,\qquad c_m^2 > 0
\label{masta}
\ee
are satisfied, there are neither ghost nor Laplacian instabilities 
in the matter sector.

For the perturbations $\psi$ and $\delta \phi$, the conditions for the absence 
of ghosts are similar to those derived in Ref.~\cite{HKT18}, i.e.,
\ba
&&
K_{11}>0
\quad {\rm or} \quad 
K_{22}>0\,,
\label{ng1}\\
&&
q_s \equiv K_{11}K_{22}-K_{12}^2 >0\,. 
\label{ng2}
\ea
Compared to the SVT theories with the action ${\cal S}_{\rm SVT}$ alone, 
the action ${\cal S}_{\rm ST}$ gives rise to modifications to  
no-ghost conditions through the change of coefficients 
$D_i$ and $w_i$, but the presence of matter 
does not affect no-ghost conditions.
Among the coefficients $D_i$ and $w_i$, there are 
the following particular relations:
\ba
&&
w_2-w_1+\tp D_6=2 H q_t\,,\\
&&
\tp^2D_1+\tp(D_4+3HD_6)-3Hw_1+w_4-w_5=3H^2q_t\,,\\
&&
w_8=3Hw_1-2w_4-\tp D_4\,,\\
&&
A_0D_8=-(2\tp D_1+D_4+3HD_6)\,.
\ea
By using these relations, one can show that the quantity $q_s$ 
is related to the determinant ${\cal D}$ 
given by Eq.~(\ref{calD}), as 
\be
q_s=\frac{H^2q_t{\cal D}}{2A_0^2(w_1-2w_2)^2}\,.
\ee
This means that, under the absence of scalar and tensor 
ghosts, the denominators in the background 
Eqs.~(\ref{dotH})-(\ref{dotA0}) do not cross 0. 

The dispersion relation in the small-scale limit is given by 
${\rm det} (c_s^2 {\bm K}-{\bm G})=0$, where $c_s$ is 
the propagation speed of scalar perturbations. 
One of the solutions is the matter propagation speed squared 
$c_m^2=G_{33}/K_{33}$, while the other two solutions are 
\ba
\hspace{-0.9cm}
c_{s1}^2
&=&
\frac{K_{11}G_{22}+K_{22}G_{11}-2K_{12}G_{12}
+\sqrt{(K_{11}G_{22}+K_{22}G_{11}-2K_{12}G_{12})^2
-4q_s(G_{11}G_{22}-G_{12}^2)}}
{2q_s},
\label{cs1}\\
\hspace{-0.9cm}
c_{s2}^2
&=&
\frac{K_{11}G_{22}+K_{22}G_{11}-2K_{12}G_{12}
-\sqrt{(K_{11}G_{22}+K_{22}G_{11}-2K_{12}G_{12})^2
-4q_s (G_{11}G_{22}-G_{12}^2)}}{2q_s}.
\label{cs2}
\ea
The Laplacian instabilities are absent 
under the conditions
\be
c_{s1}^2 > 0\,,\qquad 
c_{s2}^2 > 0\,.
\label{csscon}
\ee
Since the matrix components 
$G_{11}, G_{22}, G_{12}$ contain the 
term $\rho_m+P_m$, the existence of matter affects 
the sound speed squares $c_{s1}^2$ and $c_{s2}^2$. 
We note that there is the particular combination 
\be
q_s c_{s1}^2 c_{s2}^2=G_{11}G_{22}-G_{12}^2\,,
\label{qsre}
\ee
which is positive under the conditions 
(\ref{ng2}) and (\ref{csscon}).

The conditions (\ref{ng1}), (\ref{ng2}), and (\ref{csscon}) 
were derived under the small-scale limit. 
One may wonder what happens for the large-scale limit 
($k \to 0$). Since the Laplacian terms vanish in this limit, 
what we need to worry is the absence of scalar ghosts. 
We recall that the background Eqs.~(\ref{dotH})-(\ref{dotA0}) 
are nonsingular under the no-ghost condition $q_s>0$ 
derived for $k \to \infty$.
Since the homogenous background can be regarded as 
the $k \to 0$ limit of scalar perturbations, 
it is anticipated that the scalar perturbations in 
the large-scale limit may be stable under the condition 
$q_s>0$. However we need a separate detailed analysis 
for concrete models to support this claim, which we do not address in this paper.

\subsection{Horndeski theories}

In Horndeski theories, there are two dynamical DOFs given by 
the matrix 
\be
\vec{\mathcal{X}}^{t}=\left( \dphi, 
\frac{\delta \rho_m}{k} \right) \,,
\ee
without the perturbations $\delta A$ and $\psi$ associated 
with the vector field $A_{\mu}$.
{}From the coefficients given in Appendix~\ref{coeff}, 
there are also the following relations 
\ba
& &
D_8=D_9=D_{10}=0\,,\qquad 
D_4=-2 \dot{\phi} D_1-3H D_6\,,\nonumber \\
& &
w_2=w_3=w_5=w_6=w_7=w_8=0\,,\qquad 
w_1=\dot{\phi} D_6-2Hq_t\,,\qquad 
w_4=\dot{\phi}^2 D_1+3H \dot{\phi} D_6
-3H^2 q_t\,.
\label{correHo}
\ea
The nondynamical perturbations $\alpha, \chi, v$ obey 
Eqs.~(\ref{eqalpha}), (\ref{eqchi}), and (\ref{eqdrho}), respectively, with $w_2=w_6=w_8=0$ and ${\cal Y}=0$. 
After eliminating these fields from Eq.~(\ref{Ss}), 
the second-order scalar action can be written 
in the form (\ref{Ss2}) with the $2 \times 2$ 
matrices  ${\bm K}$, ${\bm G}$, ${\bm M}$, ${\bm B}$. 
In the small-scale limit, the nonvanishing components of 
${\bm K}$ and ${\bm G}$ are 
\ba
& &
K_{11}^{\rm Ho}
= \frac{H^2 q_t (4D_1 q_t+3D_6^2)}{w_1^2}\,,\qquad 
K_{22}^{\rm Ho}=\frac{a^2}{2(\rho_m+P_m)}\,,\\
& &
G_{11}^{\rm Ho}
= \dot{E}_2+HE_2-\frac{D_6D_7}{w_1} 
-D_2-\frac{D_6^2(\rho_m+P_m)}{2w_1^2} \,,\qquad 
G_{22}^{\rm Ho}=\frac{c_m^2 a^2}{2(\rho_m+P_m)}\,,
\label{G11Ho}
\ea
where $E_2=-D_6^2/(2w_1)$. 
We can obtain the value $K_{11}^{\rm Ho}$ by using 
$K_{22}$ in Eq.~(\ref{Kmat}) with the correspondence 
(\ref{correHo}). There is also the correspondence between 
$G_{11}^{\rm Ho}$ and $G_{22}$ in Eq.~(\ref{Gmat}), 
but we need to caution that the limit $w_3 \to 0$ 
cannot be naively taken in $G_{22}$.
Under the conditions (\ref{masta}), there are neither ghost 
nor Laplacian instabilities for the perfect fluid.

The scalar ghost associated with the perturbation $\delta \phi$ 
is absent under the condition 
\be
q_{s, {\rm Ho}} \equiv K_{11}^{\rm Ho}
=\frac{H^2q_t {\cal D}_{\rm Ho}}{w_1^2}>0\,,
\ee
where ${\cal D}_{\rm Ho}$ is defined by Eq.~(\ref{Dho}).
Provided that the scalar and tensor ghosts are absent, 
the denominators in the background Eqs.~(\ref{Ho1}) 
and (\ref{Ho2}) do not cross 0. The propagation 
speed squared of $\delta \phi$  is given by 
\be
c_{s,{\rm Ho}}^2=
\frac{G_{11}^{\rm Ho}}{q_{s, {\rm Ho}}}
=\frac{w_1^2}{H^2 q_t (4D_1 q_t+3D_6^2)} 
\left[  \dot{E}_2+HE_2-\frac{D_6D_7}{w_1} 
-D_2-\frac{D_6^2(\rho_m+P_m)}{2w_1^2}
\right]\,,
\ee
which is required to be positive to avoid the 
Laplacian instability. 
The values of $q_{s, {\rm Ho}}$ and $c_{s,{\rm Ho}}^2$ 
match with those derived in Ref.~\cite{exga}.

\subsection{Generalized Proca theories}

In generalized Proca theories, the two dynamical DOFs 
are given by 
\be
\vec{\mathcal{X}}^{t}=\left( \psi,
\frac{\delta \rho_m}{k} \right) \,,
\ee
without the scalar-field perturbation $\delta \phi$. 
In this case, we have the following relations 
\ba
D_{1,2, \cdots, 10}=0\,, \qquad w_1=w_2-2H q_t\,,\qquad 
w_4=w_5+\frac{3H(w_1+w_2)}{2}\,,\qquad 
w_8=3H w_1-2w_4\,.
\label{reGP}
\ea
Using these relations in Eqs.~(\ref{eqalpha})-(\ref{eqdrho}), eliminating $\alpha, \chi, \delta A, v$ from (\ref{Ss}), 
and taking the small-scale limit, the second-order scalar action 
reduces to the form (\ref{Ss2}) with non-vanishing 
components of  the $2\times 2$ matrices 
${\bm K}$ and ${\bm G}$: 
\ba
& &
K_{11}^{\rm GP}
= \frac{H^2 q_t (4w_5 q_t+3w_2^2)}{A_0^2 (w_1-2w_2)^2}\,,\qquad 
K_{22}^{\rm GP}=\frac{a^2}{2(\rho_m+P_m)}\,,\\
& &
G_{11}^{\rm GP}
= \dot{E_1}+H E_1-\frac{4A_0^2}{w_3}E_1^2
-\frac{w_7}{2}-\frac{w_2^2(\rho_m+P_m)}
{2A_0^2 (w_1-2w_2)^2} \,,\qquad 
G_{22}^{\rm GP}=\frac{c_m^2 a^2}{2(\rho_m+P_m)}\,,
\label{G11GP}
\ea
where $K_{11}^{\rm GP}$ and $G_{11}^{\rm GP}$ 
are the same as $K_{11}$ and $G_{11}$ in Eqs.~(\ref{Kmat}) 
and (\ref{Gmat}), respectively, with the particular relations 
(\ref{reGP}).  
The no-ghost condition for the perturbation $\psi$ is given by 
\be
q_{s,{\rm GP}} \equiv K_{11}^{\rm GP}
=\frac{H^2 q_t {\cal D}_{\rm GP}}{A_0^2 (w_1-2w_2)^2}>0\,,
\ee
where ${\cal D}_{\rm GP}$ is given by Eq.~(\ref{DGP}). 
Hence the denominators of Eqs.~(\ref{GP1}) and (\ref{GP2}) remain 
positive under the absence of scalar and tensor ghosts.
The propagation speed squared of $\psi$ yields
\be
c_{s,{\rm GP}}^2=
\frac{G_{11}^{\rm GP}}{q_{s, {\rm GP}}}
=\frac{A_0^2 (w_1-2w_2)^2}{H^2 q_t (4w_5 q_t+3w_2^2)} 
\left[  \dot{E_1}+H E_1-\frac{4A_0^2}{w_3}E_1^2
-\frac{w_7}{2}-\frac{w_2^2(\rho_m+P_m)}
{2A_0^2 (w_1-2w_2)^2}
\right]\,,
\ee
which needs to be positive for the absence of 
Laplacian instabilities. 
The values of $q_{s, {\rm GP}}$ and $c_{s,{\rm GP}}^2$ 
coincide with those obtain in Ref.~\cite{GPcosmo}.

\section{Matter perturbations and gravitational potentials}
\label{Geffsec}

In order to confront SVT theories with the observations 
associated with the evolution of matter perturbations and 
gravitational potentials, we consider non-relativistic 
matter characterized by 
\be
P_m=0\,,\qquad c_m^2=0\,.
\ee
We introduce the gauge-invariant matter density contrast 
$\delta_m$, as 
\be
\delta_m \equiv \frac{\delta \rho_m}{\rho_m}
+3H v\,.
\label{deltam}
\ee
Then, Eqs.~(\ref{eqdrho}) and (\ref{veq}) 
can be expressed as 
\ba
& &
\dot{\delta}_m -3\dot{\cal B}+\frac{k^2}{a^2} 
\left( v+\chi \right)=0\,,\label{delrhom}\\
& &
\dot{v}=\alpha\,,
\label{tv}
\ea
where ${\cal B} \equiv Hv$.
Taking the time derivative of Eq.~(\ref{delrhom}) and 
using Eq.~(\ref{tv}), we obtain 
\be
\ddot{\delta}_m+2H \dot{\delta}_m+\frac{k^2}{a^2}\Psi
=3 \left( \ddot{\cal B}+2H \dot{\cal B} \right)\,,
\label{ddotdelta}
\ee
where $\Psi$ is the gauge-invariant gravitational potential 
defined by 
\be
\Psi \equiv \alpha+\dot{\chi}\,.
\ee
We also introduce another gauge-invariant 
gravitational potential:
\be
\Phi \equiv H \chi\,,
\ee
together with the gravitational slip parameter 
\be
\eta \equiv -\frac{\Phi}{\Psi}\,.
\label{eta}
\ee
We define the effective gravitational coupling $G_{\rm eff}$ in the form 
\be
\frac{k^2}{a^2} \Psi=-4\pi \mu G \delta \rho_m\,,\quad 
{\rm with} \quad \mu=\frac{G_{\rm eff}}{G}\,.
\label{Poi}
\ee
Introducing the effective potential $\psi_{\rm eff}=\Phi-\Psi$
associated with the light bending in weak lensing 
observations \cite{lensing}, it follows that 
\be
\frac{k^2}{a^2}\psi_{\rm eff}
=8\pi G \Sigma \delta \rho_m\,,\quad 
{\rm with} \quad 
\Sigma=\frac{1+\eta}{2}\mu\,.
\label{psieff}
\ee

In what follows, we derive analytic solutions to $\Psi$, $\Phi$, $\psi$, 
and $\delta \phi$ under the quasi-static approximation 
for the modes deep inside the sound horizon ($c_s^2k^2 \gg a^2H^2$). 
This amounts to picking up terms containing $\delta \rho_m$ and 
$k^2/a^2$ in the perturbation equations 
of motion \cite{Boi,review1,DKT}. 
In some dark energy models like those in $f(R)$ gravity \cite{fR}, 
the mass of field perturbation $\delta \phi$ can be much 
lager than $H$ in the early cosmological epoch. 
In such cases, we need to take into account the mass term 
$-2D_3 \delta \phi$ in Eq.~(\ref{calZeq}). 
If the field mass is heavy, however, the scalar field hardly 
propagates, so the evolution of perturbations is similar to 
that in GR \cite{DKT}. Since we are interested in the growth 
of perturbations at the late cosmological epoch during which 
the field mass becomes as light as the Hubble expansion rate, 
it is a good approximation to neglect the masses of 
scalar and vector fields.

{}From Eq.~(\ref{delrhom}), the term $(k^2/a^2)v$ is at most of 
order $H \delta_m$. Then, for sub-horizon perturbations, we have 
$|Hv| \lesssim (aH/k)^2|\delta_m| \ll |\delta_m|$ and hence 
$\delta_m \simeq \delta \rho_m/\rho_m$ in Eq.~(\ref{deltam}). 
Moreover, the time derivatives $\ddot{\cal B}$ and $H \dot{\cal B}$ 
in Eq.~(\ref{ddotdelta}), which are at most of order 
$H^2 {\cal B}=H^3 v$ under the quasi-static approximation, 
can be neglected to the terms on its left hand side 
(which are of order $H^2 \delta_m$).
Then, from Eqs.~(\ref{ddotdelta}) and (\ref{Poi}), 
the density contrast $\delta_m$ obeys
\be
\ddot{\delta}_m+2H \dot{\delta}_m-4\pi \mu G \rho_m \delta_m 
\simeq 0\,.
\label{ddotdelta2}
\ee
After deriving analytic expressions of $\mu$ and $\Sigma$, 
we can solve Eqs.~(\ref{ddotdelta2}), (\ref{Poi}), and 
(\ref{psieff})  for $\delta_m$, $\Psi$, and 
$\psi_{\rm eff}$, respectively.

\subsection{Full SVT theories}

We derive analytic solutions to $\mu$ and $\Sigma$ in full SVT 
theories with the three dynamical scalar perturbations
$\psi, \delta \phi$, and $\delta \rho_m$.
Applying the quasi-static approximation to 
Eqs.~(\ref{eqalpha}) and (\ref{eqdA}), we have
\ba
\delta \rho_m 
&\simeq& \frac{k^2}{a^2} \left( w_6 \psi-w_1 \chi
-D_6 \delta \phi-{\cal Y} \right)\,,\\ 
{\cal Y}
&\simeq& \left( w_6 -\frac{w_2}{A_0} \right)\psi-2w_2 \chi\,,
\label{Yeq2}
\ea
and hence 
\be
\delta \rho_m \simeq
-\frac{k^2}{a^2} \left( \frac{w_1-2w_2}{H} \Phi
-\frac{w_2}{A_0}\psi+D_6 \delta \phi \right)\,.
\label{quasi1}
\ee
Eliminating the perturbation $v$ from Eqs.~(\ref{eqchi}) 
and (\ref{eqdrho}), it follows that 
\be
\dot{\delta \rho}_m+3H \delta \rho_m
+\frac{k^2}{a^2} \left( \rho_m \chi-w_1 \alpha
+w_2 \frac{\delta A}{A_0}+D_6 \dot{\delta \phi}
-D_7 \delta \phi \right)=0\,.
\label{delrhom2}
\ee
Substituting Eq.~(\ref{quasi1}) and its time derivative 
into Eq.~(\ref{delrhom2}), the derivative 
term $\dot{\delta \phi}$ cancels out. 
After this substitution the time derivative $\dot{\psi}$ 
appears, but it can be eliminated by using the relation 
\be
\dot{\psi} \simeq 2A_0 \alpha-\delta A
+\frac{1}{w_3} \left[ (w_2-A_0 w_6)\psi
+2w_2 A_0 \chi \right]\,, 
\ee
which follows from Eqs.~(\ref{Ydef}) and (\ref{Yeq2}). 
Then, we obtain 
\be
w_3 A_0^2 \left( \kappa_1 \Psi+\kappa_2 \Phi
+\kappa_4 \delta \phi \right)
+\kappa_3 \psi  \simeq 0\,,
\label{quasi2}
\ee
where 
\ba
\kappa_1 &=& w_1-2w_2\,,\\
\kappa_2 &=& 
\frac{1}{H} \left( \dot{\kappa}_1+H\kappa_1
-\rho_m-\frac{2w_2^2}{w_3} \right)\,,\\
\kappa_3 &=& 
w_2 w_6 A_0^2-\left( \dot{w}_2 w_3+Hw_2 w_3+w_2^2 
\right)A_0+w_2w_3 \dot{A}_0\,,\label{ka3}\\
\kappa_4 &=& 
\dot{D}_6+HD_6+D_7\,.
\ea
We also substitute Eq.~(\ref{Yeq2}) and its time derivative 
into Eq.~(\ref{calYeq}). This leads to 
\be
2w_2 w_3 A_0^2 \Psi-\frac{2A_0}{H}\kappa_3\Phi
+\kappa_5 \psi-2D_{10}w_3A_0^3 \delta \phi=0\,,
\label{quasi3}
\ee
where 
\be
\kappa_5=\left( 2w_3 w_7+w_6^2 \right)A_0^3
-\left[ (\dot{w}_6+Hw_6)w_3+2w_2w_6 \right]A_0^2
+\left[ (\dot{w}_2+Hw_2+w_6 \dot{A}_0)w_3+w_2^2 
\right]A_0-2w_2w_3 \dot{A}_0\,.\label{ka5}
\ee
{}From Eqs.~(\ref{calZeq}) and (\ref{Zdef}), it follows that 
\be
HD_6 \Psi+\kappa_4 \Phi+HD_{10}\psi
-2HD_2 \delta \phi \simeq 0\,.
\label{quasi4}
\ee
Now, we can solve Eqs.~(\ref{quasi1}), (\ref{quasi2}), (\ref{quasi3}), 
and (\ref{quasi4}) for $\Psi, \Phi, \psi$, and $\delta \phi$, as
\ba
\hspace{-0.9cm}
\Psi &\simeq& 
-\frac{4A_0^8D_{10}^2H \kappa_2 q_v^2
+4A_0^4 D_{10}\kappa_3 \kappa_4 q_v
+2A_0^3D_2 H \kappa_2 \kappa_5 q_v
+A_0^3 \kappa_4^2 \kappa_5 q_v
-2D_2 \kappa_3^2}{\Delta}
\frac{a^2}{k^2} \delta \rho_m\,,\label{Psiq}\\
\hspace{-0.9cm}
\Phi &\simeq& 
\frac{A_0^3 H q_v (4A_0^5 D_{10}^2 \kappa_1 q_v
+4A_0^4 D_{10}\kappa_4 w_2 q_v+2A_0D_6 D_{10}
\kappa_3+2D_2 \kappa_1 \kappa_5-4D_2\kappa_3 w_2
+D_6 \kappa_4 \kappa_5)}{\Delta}
\frac{a^2}{k^2} \delta \rho_m\,,\\
\hspace{-0.9cm}
\psi &\simeq&
\frac{2A_0^4 q_v (2A_0^4 D_6 D_{10}H \kappa_2 q_v
-2A_0^4 D_{10} \kappa_1 \kappa_4 q_v-4A_0^3 
D_2 H \kappa_2 w_2 q_v-2A_0^3 \kappa_4^2 w_2 q_v
+2D_2\kappa_1 \kappa_3+D_6 \kappa_3 \kappa_4)}{\Delta}
\frac{a^2}{k^2} \delta \rho_m\,,\\
\hspace{-0.9cm}
\delta \phi &\simeq&
-\frac{4A_0^7D_{10}H \kappa_2 w_2 q_v^2
-2A_0^4 D_{10}\kappa_1 \kappa_3 q_v
+A_0^3 D_6 H \kappa_2 \kappa_5 q_v
-A_0^3 \kappa_1 \kappa_4 \kappa_5 q_v
+2A_0^3 \kappa_3 \kappa_4 w_2 q_v-D_6\kappa_3^2}{\Delta}
\frac{a^2}{k^2} \delta \rho_m\,, \label{delphiq}
\ea
where we used the relation $w_3=-2A_0^2q_v$, and $\Delta$ 
is defined by 
\ba
\Delta &=& -4 A_0^8 D_{10}^2 \kappa_1^2 q_v^2
+8A_0^7 D_{10}w_2 q_v^2 (D_6 H \kappa_2 
-\kappa_1 \kappa_4)
-4A_0^6 w_2^2 q_v^2 (2H D_2 \kappa_2+\kappa_4^2) 
-4A_0^4 D_6 D_{10} \kappa_1 \kappa_3 q_v
\nonumber \\
& &
+A_0^3 q_v \left[ D_6(D_6H \kappa_2 \kappa_5
+4\kappa_3 \kappa_4 w_2)
+2\kappa_1(4D_2 \kappa_3 w_2-D_6\kappa_4 \kappa_5) 
-2D_2 \kappa_1^2 \kappa_5 \right]
-D_6^2 \kappa_3^2\,.
\label{Delta1}
\ea
We compute the right hand side of Eq.~(\ref{qsre}) 
by using the definitions (\ref{Gmat}) with Eq.~(\ref{E123}). 
Then, the determinant $\Delta$ is simply related to the quantity 
$q_s c_{s1}^2 c_{s2}^2$, as 
\be
\Delta=16 \kappa_1^2 A_0^8 q_v^2 
q_s c_{s1}^2 c_{s2}^2\,, 
\label{Delta2}
\ee
which is positive under the absence of ghost and 
Laplacian instabilities of scalar perturbations.

{}From Eqs.~(\ref{eta})-(\ref{psieff}), 
the quantities $\mu$ and $\Sigma$ can be estimated as 
\ba
\mu
&=& \frac{4A_0^8D_{10}^2H \kappa_2 q_v^2
+4A_0^4 D_{10}\kappa_3 \kappa_4 q_v
+A_0^3 \kappa_5 q_v (2D_2 H \kappa_2+\kappa_4^2)
-2D_2 \kappa_3^2}{64\pi G \kappa_1^2 A_0^8 q_v^2 
q_s c_{s1}^2 c_{s2}^2}\,,\label{muf}\\
\Sigma
&=&\frac{1}{128\pi G \kappa_1^2 A_0^8 q_v^2 
q_s c_{s1}^2 c_{s2}^2} 
\left[ 4 A_0^8 D_{10}^2 H q_v^2 (\kappa_1+\kappa_2)
+4A_0^7 D_{10} H \kappa_4 w_2 q_v^2
+2A_0^4 D_{10} \kappa_3 q_v (D_6 H+2\kappa_4) 
\right.
\nonumber \\
& &\left.+A_0^3 q_v (2D_2 H \kappa_1 \kappa_5
+2D_2 H \kappa_2 \kappa_5 -4D_2 H \kappa_3 w_2 
+D_6 H \kappa_4 \kappa_5 +\kappa_4^2 \kappa_5)
-2D_2 \kappa_3^2 \right]\,.
\label{Sigmaf}
\ea
The two quantities $\mu$ and $\Sigma$ evolve differently 
depending on the models of dark energy. 
Since the evolution of gravitational potentials as well as 
the growth of matter perturbations is affected by the changes 
of $\mu$ and $\Sigma$, one can distinguish between dark energy models 
in SVT theories from the observation data of large-scale structures, 
weak lensing, and CMB \cite{obcon}.

\subsection{Horndeski theories}

In Horndeski theories, the scalar perturbation $\psi$ does not 
exist as a propagating DOF. Under the quasi-static approximation, 
we have three perturbation equations of motion: (\ref{quasi1}) with $w_2=0$, 
(\ref{quasi4}) with $D_{10}=0$, and 
\be
w_1 \Psi+\kappa_2 \Phi+\kappa_4 \delta \phi
\simeq 0\,,
\ee
where the last equation is the analogue of Eq.~(\ref{quasi2}).
Solving these equations for $\Psi, \Phi, \delta \phi$, 
it follows that 
\be
\Psi \simeq -\frac{2D_2 H \kappa_2+\kappa_4^2}
{\Delta_{\rm Ho}} \frac{a^2}{k^2} \delta \rho_m\,,\qquad 
\Phi \simeq \frac{H(2D_2  w_1+D_6 \kappa_4)}
{\Delta_{\rm Ho}} \frac{a^2}{k^2} \delta \rho_m\,,\qquad 
\delta \phi \simeq  -\frac{D_6 H \kappa_2-w_1\kappa_4}
{\Delta_{\rm Ho}} \frac{a^2}{k^2} \delta \rho_m\,,
\ee
where 
\be
\Delta_{\rm Ho}
=D_6^2 H \kappa_2-2D_2 w_1^2-2D_6 w_1 \kappa_4\,.
\ee
The determinant $\Delta_{\rm Ho}$ is related to the quantity 
$G_{11}^{\rm Ho}=q_{s,{\rm Ho}}c_{s,{\rm Ho}}^2$ in 
Eq.~(\ref{G11Ho}), as
\be
\Delta_{\rm Ho}=2w_1^2\,q_{s,{\rm Ho}}c_{s,{\rm Ho}}^2\,.
\ee
Then, the quantities $\mu$ and $\Sigma$ 
reduce, respectively, to 
\be
\mu=\frac{2D_2 H \kappa_2+\kappa_4^2}
{8\pi G w_1^2  q_{s,{\rm Ho}}c_{s,{\rm Ho}}^2}\,,
\qquad 
\Sigma=\frac{2D_2 H (\kappa_1+\kappa_2)
+D_6 H\kappa_4+\kappa_4^2}
{16 \pi G  w_1^2  q_{s,{\rm Ho}}c_{s,{\rm Ho}}^2}\,.
\ee
We confirmed that these results agree with 
those derived in Ref.~\cite{DKT} in the limit that 
the scalar-field mass vanishes.

\subsection{Generalized Proca theories}

In generalized Proca theories the scalar-field perturbation 
$\delta \phi$ is absent, so there are three independent 
perturbation equations: (\ref{quasi1}) with $D_6=0$, 
(\ref{quasi2}) with $\kappa_4=0$, and 
(\ref{quasi3}) with $D_{10}=0$.
Solving these equations for $\Psi, \Phi, \psi$, we obtain
\be
\Psi \simeq -\frac{A_0 H \kappa_2 \kappa_5 w_3
+2\kappa_3^2}{\Delta_{\rm GP}} \frac{a^2}{k^2} \delta \rho_m,\quad 
\Phi \simeq \frac{A_0 w_3 H
(\kappa_1 \kappa_5-2\kappa_3 w_2)}
{\Delta_{\rm GP}} \frac{a^2}{k^2} \delta \rho_m,\quad 
\psi \simeq  \frac{2A_0^2 w_3 (A_0 H \kappa_2 w_2 w_3+\kappa_1 \kappa_3)}
{\Delta_{\rm GP}} \frac{a^2}{k^2} \delta \rho_m\,,
\ee
where 
\be
\Delta_{\rm GP}
=A_0 w_3 \left(2A_0 H \kappa_2 w_2^2 w_3-\kappa_1^2 \kappa_5
+4\kappa_1 \kappa_3 w_2 \right)\,.
\ee
The determinant $\Delta_{\rm GP}$ can be expressed 
in terms of the quantity $G_{11}^{\rm GP}
=q_{s,{\rm GP}}\,c_{s,{\rm GP}}^2$ in 
Eq.~(\ref{G11GP}), as
\be
\Delta_{\rm GP}=16A_0^8 \kappa_1^2 
q_v^2\,q_{s,{\rm GP}}\,c_{s,{\rm GP}}^2\,.
\ee
On using this relation, it follows that 
\be
\mu=\frac{\kappa_3^2-A_0^3 H \kappa_2 \kappa_5 
q_v}{32\pi  G A_0^8 \kappa_1^2 q_v^2\, 
q_{s,{\rm GP}}\,c_{s,{\rm GP}}^2}\,,\qquad 
\Sigma=\frac{\kappa_3^2-A_0^3 H 
q_v [\kappa_5(\kappa_1+\kappa_2)-2\kappa_3 w_2]}
{64\pi G A_0^8 \kappa_1^2 
q_v^2\,q_{s,{\rm GP}}\,c_{s,{\rm GP}}^2}\,,
\ee
which coincide with those derived in Ref.~\cite{Geff16}.

\section{SVT theories with $c_t^2=1$}
\label{conmodelsec}

Finally, we estimate the quantities 
$\mu$ and $\Sigma$ associated with Newtonian and weak lensing 
potentials for SVT theories in which $c_t^2$ is exactly 
equivalent to 1. We consider the couplings of the forms
\be
G_4=G_4(\phi)\,,\qquad G_5=0\,, \qquad 
f_4=0\,,\qquad f_5=0\,,
\label{G45}
\ee
where the $\phi$ dependence of $f_4$ has been 
absorbed into $G_4(\phi)$. 
We take into account all the Lagrangians associated 
with intrinsic vector modes, which affect scalar perturbations 
only through the quantity 
\be
w_3=-2A_0^2 q_v\,.
\ee
The variables $w_1, w_2, w_6, w_7$, which appear 
in $\kappa_1, \kappa_2, \kappa_3, \kappa_5$, are given by 
\ba
w_1 &=& -G_{3,X_1} \dot{\phi}^3 -2 G_{4,\phi} \dot{\phi}
+2A_0^3 \left( f_{3,X_3}+\tilde{f}_3 \right)-4G_4H\,,\label{w1} \\
w_2 &=&-A_0 w_6= 2A_0^3  \left( f_{3,X_3}+\tilde{f}_3 \right)\,,\\
w_7 &=& 2\dot{A}_0 \left( f_{3,X_3}+\tilde{f}_3 \right)
+\frac{\tp(f_{2,X_2}+4f_{3,\phi})}{2A_0}\,.\label{w7}
\ea
In Eqs.~(\ref{muf}) and (\ref{Sigmaf}) there are also other variables 
$\kappa_4, D_2, D_6, D_{10}$, whose explicit forms are 
\ba
\kappa_4 &=&
f_{2,X_1} \dot{\phi}+\frac12 f_{2,X_2}A_0
+\dot{\phi} \left( 2G_{3,\phi}-2\ddot{\phi} G_{3,X_1}
-4G_{3,X_1}H \dot{\phi}-G_{3,X_1 \phi} \dot{\phi}^2
-G_{3,X_1X_1}\dot{\phi}^2 \ddot{\phi} \right) \nonumber \\
& &+2f_{3,\phi}A_0-4H G_{4,\phi}+2f_{4,\phi \phi} \dot{\phi}\,,
\label{ka4} \\
D_2 &=& 
-\frac{1}{2}f_{2,X_1}-G_{3,\phi}+G_{3,X_1}\ddot{\phi} 
+\frac{1}{2} \dot{\phi} \left( 4G_{3,X_1}H+G_{3,X_1 \phi} 
\dot{\phi}+G_{3,X_1 X_1}\dot{\phi} \ddot{\phi} 
\right)\,,\\ 
D_6 &=& -G_{3,X_1}\dot{\phi}^2 -2G_{4,\phi}\,,\\
D_{10} &=& \frac{1}{2} f_{2,X_2}+2f_{3,\phi}\,.
\label{D10}
\ea
Note that $\mu$ and $\Sigma$ also depend on 
the combination $q_s c_{s1}^2 c_{s2}^2$, 
which can be expressed in terms of the variables 
mentioned above as well as $q_v$. 
Unless we impose further conditions, none of the 
variables given above vanish.
As we see in Eqs.~(\ref{muf}) and (\ref{Sigmaf}), the effect 
of intrinsic vector modes on $\mu$ and $\Sigma$ is 
generally present through the quantity $q_v$ even for 
the theories with $c_t^2=1$.

There are classes of SVT theories with $c_t^2=1$ in which 
the dependence of $q_v$ in $\mu$ and $\Sigma$ disappears. 
In the following, we focus on the theories satisfying
\be
f_3=f_3(\phi)\,,\qquad 
\tilde{f}_3=0\,,
\label{f3}
\ee
besides the conditions (\ref{G45}).  
In this case, the variables (\ref{w1})-(\ref{w7}) reduce, respectively, to 
\be
w_1= -G_{3,X_1}\dot{\phi}^3 
-2G_{4,\phi} \dot{\phi}-4G_4H\,,
\qquad 
w_2=w_6=0\,,
\qquad 
w_7=-f_{2,X_3}\,.
\label{w127}
\ee
For the derivation of $w_7$, we exploited the fact that 
the background Eq.~(\ref{back4}) gives 
\be
\left( f_{2,X_2}+4f_{3,\phi} \right) \dot{\phi}=-2A_0f_{2,X_3}\,.
\label{backco}
\ee
On using the relations (\ref{w127}) in Eqs.~(\ref{ka3}) 
and (\ref{ka5}), it follows that 
\be
\kappa_3=0\,,\qquad 
\kappa_5=2w_3w_7 A_0^3=4A_0^5q_v f_{2,X_3}\,.
\ee
Then, the terms containing $q_v^2$ in the denominators 
and numerators of Eqs.~(\ref{muf}) and (\ref{Sigmaf}) 
are factored out, such that 
\ba
\mu &=& \frac{f_{2,X_3} (2D_2 H \kappa_2+\kappa_4^2)
+D_{10}^2 H \kappa_2}
{16\pi G \kappa_1^2 q_s c_{s1}^2 c_{s2}^2}\,,\label{muex}\\
\Sigma &=& \frac{f_{2,X_3} [ H(2D_2 \kappa_1
+2D_2 \kappa_2+D_6 \kappa_4)+\kappa_4^2]
+D_{10}^2 H (\kappa_1+\kappa_2)}
{32\pi G \kappa_1^2 q_s c_{s1}^2 c_{s2}^2}\,.
\label{Sigmaex}
\ea
Now, we substitute the values $\kappa_1=w_1$, 
$\kappa_2=(\dot{w}_1+Hw_1-\rho_m)/H$, and 
$\kappa_4$ into Eqs.~(\ref{muex}) and (\ref{Sigmaex}).
In doing so, we take the time derivative of 
$w_1=-G_{3,X_1}\dot{\phi}^3-2 G_{4,\phi}\dot{\phi}-4G_4H$ 
and then use the background Eq.~(\ref{back2}), i.e., 
\be
4G_4 \dot{H}+\left( G_{3,X_1}\dot{\phi}^2 +2G_{4,\phi} 
\right) \ddot{\phi}+\left( 
f_{2,X_1} \dot{\phi}+2G_{3,\phi} \dot{\phi} 
-3G_{3,X_1}H \dot{\phi}^2-2G_{4,\phi}H
+2G_{4,\phi \phi} \dot{\phi} \right)\dot{\phi}
-A_0^2 f_{2,X_3}=-\rho_m-P_m\,,
\ee
to eliminate $\rho_m$ in $\kappa_2$ (with $P_m=0$). 
Then, we obtain 
\ba
\mu &=& \frac{1}{16\pi G\,G_4} \left[ 1
+\frac{(G_{3,X_1}\dot{\phi}^2
-2G_{4,\phi})^2}{\xi_s} 
\right]\,,\label{mumo1}\\
\Sigma &=&
\frac{1}{16\pi G\,G_4} \left[ 1
+\frac{G_{3,X_1}\dot{\phi}^2(G_{3,X_1}\dot{\phi}^2
-2G_{4,\phi})}{\xi_s} \right]\,,
\label{mumo2}
\ea
where 
\be
\xi_s \equiv 
\frac{w_1^2 q_s c_{s1}^2 c_{s2}^2}{f_{2,X_3}G_4 H^2}
=\frac{2A_0}{\dot{\phi}} 
\left( f_{2,X_2}+4f_{3,\phi} \right) G_4+\xi_{\rm Ho}\,,
\label{w1q}
\ee
with 
\be
\xi_{\rm Ho} \equiv 4G_4 \left[ f_{2,X_1}+2G_{3,\phi}-2G_{3,X_1} 
\left( \ddot{\phi}+2 H \dot{\phi} \right) 
-\dot{\phi}^2 \left( G_{3,X_1 \phi}+G_{3,X_1 X_1} 
\ddot{\phi} \right) \right]
-G_{3,X_1}^2 \dot{\phi}^4
+4G_{4,\phi} \left( G_{3,X_1} \dot{\phi}^2
+3 G_{4,\phi} \right).
\ee

{}From Eqs.~(\ref{mumo1}) and (\ref{mumo2}), we observe that 
the $X_1$ dependence in $G_3$ and the $\phi$ dependence in 
$G_4$ lead to modifications to the values of $\mu$ and 
$\Sigma$ in GR. Since the two conditions $q_t=2G_4>0$ and 
$q_s c_{s1}^2 c_{s2}^2>0$ are required for the absence of ghost 
and Laplacian instabilities, the second term in the square bracket 
of Eq.~(\ref{mumo1}) is either positive or negative 
depending on the sign of $f_{2,X_3}$. 
If $f_{2,X_3}>0$, then the gravitational attraction is enhanced 
by the couplings $G_3(X_1)$ and $G_4(\phi)$. 
{}From Eq.~(\ref{ddotdelta2}), the growth of $\delta_m$ 
is also modified from that in GR.

A subclass of Horndeski theories allows the existence of 
no slip gravity scenario with $c_t^2=1$ 
in which the effective gravitational couplings to 
matter and light are equivalent to each other \cite{Linder18}, 
such that $\mu=\Sigma$. 
In SVT theories given by the functions (\ref{f3}), 
it follows from Eqs.~(\ref{mumo1}) and (\ref{mumo2}) 
that the condition for no slip gravity corresponds to 
$G_{4,\phi}=0$ or $G_{3,X_1}\dot{\phi}^2=2G_{4,\phi}$.
In the former case the gravitational interactions are enhanced 
by the coupling $G_{3,X_1}$, while the enhancement 
is absent for the latter case.

For the theories in which the conditions 
\be
f_{2,X_2}=0\,,\quad {\rm and} \quad  f_{3,\phi}=0
\label{f2X2}
\ee
are satisfied, the term $\xi_s$ in Eq.~(\ref{w1q}) 
is equivalent to $\xi_{\rm Ho}$.
Then, Eqs.~(\ref{mumo1}) and (\ref{mumo2}) 
reduce to
\ba
\mu &=& \frac{1}{16\pi G\,G_4} \left[ 1
+\frac{(G_{3,X_1}\dot{\phi}^2
-2G_{4,\phi})^2}{\xi_{\rm Ho}} 
\right]\,,\label{mumo1d}\\
\Sigma &=&
\frac{1}{16\pi G\,G_4} \left[ 1
+\frac{G_{3,X_1}\dot{\phi}^2(G_{3,X_1}\dot{\phi}^2
-2G_{4,\phi})}{\xi_{\rm Ho}} 
\right]\,.\label{mumo2d}
\ea
These values are equivalent to those in the subclass of 
Horndeski theories given by the Lagrangian 
${\cal L}=f_2(\phi,X_1)+G_3(\phi,X_1) \square \phi
+G_4(\phi) R$ \cite{DKT}. 
This means that the effect of the vector field on $\mu$ 
and $\Sigma$ arises from the kinetic mixing 
between $A_0$ and $\dot{\phi}$ (i.e., $f_{2,X_2} \neq 0$) 
as well as from the cubic scalar-vector 
coupling $f_3(\phi)g^{\mu \nu}S_{\mu \nu}$.
As long as the condition $f_{2,X_2}+4f_{3,\phi} \neq 0$ 
is satisfied,  Eq.~(\ref{backco}) shows that the temporal vector 
component $A_0$ does not vanish for $f_{2,X_3} \neq 0$. 
In this case, $\mu$ and $\Sigma$ in Eqs.~(\ref{mumo1}) 
and (\ref{mumo2}) depend on both the scalar-vector 
couplings $f_{2,X_2}+4f_{3,\phi}$ and 
the contribution $\xi_{\rm Ho}$ arising in Horndeski theories.

Finally, let us consider the theories satisfying 
\be
G_3=0\,, 
\ee
without imposing the condition (\ref{f2X2}). 
Then, Eqs.~(\ref{mumo1}) and (\ref{mumo2}) yield
\ba
\mu &=& \frac{1}{16\pi G\,G_4} \left[ 
1+\frac{2G_{4,\phi}^2}
{(A_0/\dot{\phi})(f_{2,X_2}+4f_{3,\phi})G_4
+2(G_4f_{2,X_1}+3G_{4,\phi}^2)} 
\right]\,,\label{muG3}\\
\Sigma &=&   \frac{1}{16\pi G\,G_4}\,.
\ea
Provided that $f_{2,X_2}+4f_{3,\phi} \neq 0$, 
the scalar-vector interactions
give rise to the modification to $\mu$, 
while $\Sigma$ is not affected.
Taking the limit $f_{2,X_2}+4f_{3,\phi} \to 0$ 
in Eq.~(\ref{muG3}),  
we recover the value of $\mu$ derived for a nonminimally 
coupled scalar field with the Lagrangian 
${\cal L}=f_2(\phi,X_1)+G_4(\phi) R$ \cite{Boi,ST07}.

The fact that $\mu$ and $\Sigma$ are independent of $q_v$ 
is attributed to the choice of functions $f_3$ and 
$\tilde{f}_3$ in Eq.~(\ref{f3}). 
For the theories in which $f_3$ depends on $X_3$ and 
$\tilde{f}_3$ is a non-vanishing function, 
the quantity $\kappa_3$ does not vanish 
in Eqs.~(\ref{muf}) and (\ref{Sigmaf}) and 
hence $\mu$ and $\Sigma$ 
depend on $q_v$. In such cases, the effect of 
the vector field on $\mu$ and $\Sigma$ 
arises not only through scalar-vector interactions
but also from intrinsic vector modes.

\section{Conclusions}
\label{concludesec}

In full SVT theories with second-order equations of motion and parity invariance, we studied the 
behavior of linear cosmological perturbations on the flat 
FLRW background. 
The difference from previous study \cite{HKT18} is 
that we have taken into account the perfect fluid 
and the Horndeski action (\ref{Hoaction}) besides 
the SVT action (\ref{action}). 
As a result, the perturbation equations of motion can be 
directly applied to the growth of matter perturbations and 
the evolution of gravitational potentials for dark energy models 
in the framework of SVT theories.
Moreover, our general analysis can 
accommodate both Horndeski and generalized 
Proca theories as subclasses of the action (\ref{actionfull}). 

In Sec.~\ref{backsec}, we derived the background 
equations of motion with the matter 
density $\rho_m$ and pressure $P_m$. 
In particular, Eqs.~(\ref{back2}), (\ref{back3}) and 
the time derivative of  (\ref{back4}) are 
expressed in compact forms by using the coefficients 
present in the 
second-order action of scalar perturbations.
We showed that, for a nonvanishing determinant 
${\cal D}$,  the dynamical system in SVT theories can 
be solved for $\dot{H}, \ddot{\phi}, \dot{A}_0$ 
in the forms (\ref{dotH})-(\ref{dotA0}). 
In Horndeski and generalized Proca theories, 
the dynamical systems are described by 
the combinations
$(\dot{H}, \ddot{\phi})$ and $(\dot{H}, \dot{A}_0)$, 
respectively. 
In all cases, the determinant ${\cal D}$ 
is directly related to a quantity $q_s$ associated 
with the no-ghost condition of scalar perturbations.

In Sec.~\ref{TVsec}, we decomposed the perturbations 
of metric, scalar and vector fields, and perfect fluid into 
tensor, vector, and scalar modes and obtained 
the second-order actions of tensor and vector perturbations. 
The conditions for the absence of ghost and Laplacian 
instabilities in the tensor sector correspond 
to Eq.~(\ref{qtcon}), with the tensor propagation 
speed squared given by Eq.~(\ref{ct}). 
If we apply SVT theories to dark energy and strictly demand that $c_t^2=1$ without allowing any tuning 
among functions, the theories are restricted to be of 
the forms (\ref{ctcon}). 
For vector perturbations, we showed that neither 
the scalar-tensor action ${\cal S}_{\rm ST}$ nor 
the perfect-fluid action ${\cal S}_{m}$ give rise to 
modifications to stability conditions in the small-scale 
limit derived in Ref.~\cite{HKT18}.

In Sec.~\ref{scasec}, we derived the second-order action of 
scalar perturbations and the resulting full perturbation 
equations of motion in the scalar sector.
In SVT theories, there are three scalar dynamical DOFs 
$\psi, \delta \phi, \delta \rho_m$, among which the matter 
perturbation $\delta \rho_m$ is decoupled from others 
in the small-scale limit. 
The no-ghost conditions for the perturbations $\psi$ and 
$\delta \phi$ correspond to Eqs.~(\ref{ng1})-(\ref{ng2}), 
while their propagation speed squares are given by 
Eqs.~(\ref{cs1})-(\ref{cs2}). 
The perfect fluid affects the values of $c_{s1}^2$ and 
$c_{s2}^2$ through the term $\rho_m+P_m$ appearing 
in $G_{11}, G_{22}, G_{12}$ of Eq.~(\ref{Gmat}). 
We also showed that our general framework recovers 
the stability conditions of scalar perturbations
derived in Horndeski and generalized Proca theories.

In Sec.~\ref{Geffsec}, we studied the behavior of matter 
perturbations $\delta \rho_m$ and gauge-invariant 
gravitational potentials $\Psi,\Phi$ by employing the 
quasi-static approximation for scalar perturbations deep 
inside the sound horizon. Under this approximation 
scheme, we derived the closed-form expressions of 
$\Psi, \Phi, \psi, \delta \phi$ in SVT theories in the forms 
(\ref{Psiq})-(\ref{delphiq}), where the determinant 
$\Delta$ in denominators can be expressed 
in terms of the quantity $q_s c_{s1}^2 c_{s2}^2$, 
as Eq.~(\ref{Delta2}). 
The dimensionless quantities $\mu$ and $\Sigma$ 
associated with Newtonian and weak lensing gravitational 
potentials $\Psi$ and $\psi_{\rm eff}$ are given by 
Eqs.~(\ref{muf}) and (\ref{Sigmaf}), respectively. 
They contain quantities like $q_v$ and 
$q_s c_{s1}^2 c_{s2}^2$, whose positivities are 
required for the stability of 
vector and scalar perturbations.
We also reproduced the values of $\mu$ and $\Sigma$ 
in Horndeski and generalized Proca theories 
as specific cases.

In Sec.~\ref{conmodelsec}, we applied our general formulas 
of $\mu$ and $\Sigma$ to the SVT theories 
satisfying $c_t^2=1$. 
If the cubic couplings $f_3(X_3)$ and $\tilde{f}_3$ 
are present, the intrinsic vector modes generally affect 
$\mu$ and $\Sigma$ through the quantity $q_v$. 
For the theories with $f_3=f_3(\phi)$ and $\tilde{f}_3=0$, 
we found that the terms containing $q_v^2$ in Eqs.~(\ref{muf}) 
and (\ref{Sigmaf}) are factored out. 
In latter theories, the evolution of scalar perturbations 
for the modes relevant to the observations of large-scale 
structures and weak lensing is not affected by intrinsic 
vector modes. However, as long as the scalar-vector couplings 
$f_{2,X_2}$ and $f_{3,\phi}$ are present, their effects appear 
as the combination $f_{2,X_2}+4f_{3,\phi}$
in the expressions of 
$\mu$ and $\Sigma$ given by Eqs.~(\ref{mumo1}) and 
(\ref{mumo2}). In such cases, the behavior 
of matter perturbations and gravitational potentials 
is modified by the scalar-vector interactions. 
 
Our general results about the stabilities of perturbations 
can be directly applied to the construction of viable dark energy models in the framework 
of SVT theories. Moreover, for the models with $c_t^2=1$, it will be of interest to study their observational signatures in more detail to extract some new features in 
SVT theories. These issues are left for future works.

\section*{Acknowledgements}

We thank Lavinia Heisenberg for useful discussions. 
RK is supported by the Grant-in-Aid for Young 
Scientists B of the JSPS No.\,17K14297.  
ST is supported by the Grant-in-Aid 
for Scientific Research Fund of the JSPS No.~16K05359 and 
MEXT KAKENHI Grant-in-Aid for 
Scientific Research on Innovative Areas ``Cosmic Acceleration'' (No.\,15H05890). 

\appendix

%
\section{Coefficients in the second-order action of scalar perturbations}
\label{coeff}

The coefficients $D_{1,\cdots,10}$ and $w_{1,\cdots,8}$ 
appearing in the background Eqs.~(\ref{back2}), (\ref{back3}),  
(\ref{back5}) and the second-order action of 
scalar perturbations (\ref{Ss}) are given by
\ba
&&
D_1=H^3\tp\left(3 G_{5,X_1}+\frac72\tp^2G_{5,X_1X_1}+\frac12\tp^4G_{5,X_1X_1X_1}\right)
+3H^2\left[G_{4,X_1}-G_{5,\phi}+\tp^2\left(4G_{4,X_1X_1}-\frac52G_{5,X_1\phi}\right)\right.
\notag\\
&&\hspace{0.9cm}
\left.+\tp^4\left(G_{4,X_1X_1X_1}-\frac12G_{5,X_1X_1\phi}\right)\right]
-3H\tp\left[G_{3,X_1}+3G_{4,X_1\phi}+\tp^2\left(\frac12G_{3,X_1X_1}+G_{4,X_1X_1\phi}\right)\right]
\notag\\
&&\hspace{0.9cm}
+\frac{1}{2}\left[f_{2,X_1}+2G_{3,\phi}+\tp^2\left(f_{2,X_1X_1}+G_{3,X_1\phi}\right)
+\tp A_0 f_{2,X_1X_2}+\frac{A_0^2}{4}f_{2,X_2X_2}\right]
\,,
\notag\\
&&
D_2=-\left[2(G_{4,X_1}-G_{5,\phi})+\tp^2(2G_{4,X_1X_1}-G_{5,X_1\phi})
+H\tp(2G_{5,X_1}+\tp^2G_{5,X_1X_1})\right]\dot{H}
\notag\\
&&\hspace{0.9cm}
+\left[G_{3,X_1}+3G_{4,X_1\phi}+\tp^2\left(\frac{G_{3,X_1X_1}}{2}+G_{4,X_1X_1\phi}\right)
-2H\tp(3G_{4,X_1X_1}-2G_{5,X_1\phi})\right.
\notag\\
&&\hspace{0.9cm}
\left.-H\tp^3(2G_{4,X_1X_1X_1}-G_{5,X_1X_1\phi})
-H^2\left(G_{5,X_1}+\frac52\tp^2G_{5,X_1X_1}+\frac12\tp^4G_{5,X_1X_1X_1}
\right)
\right]\ddot{\phi}
\notag\\
&&\hspace{0.9cm}
-H^3\tp\left(2G_{5,X_1}+\tp^2G_{5,X_1X_1}\right)
-H^2\left[3(G_{4,X_1}-G_{5,\phi})+5\tp^2\left(G_{4,X_1X_1}-\frac12G_{5,X_1\phi}\right)
+\frac12\tp^4G_{5,X_1X_1\phi}\right]
\notag\\
&&\hspace{0.9cm}
+2H\tp(G_{3,X_1}+3G_{4,X_1\phi})-H\tp^3(2G_{4,X_1X_1\phi}-G_{5,X_1\phi\phi})
+\tp^2\left(\frac12G_{3,X_1\phi}+G_{4,X_1\phi\phi}\right)-G_{3,\phi}-\frac12 f_{2,X_1}\,,
\notag\\
&&
D_3=3\left[G_{4,\phi\phi}+f_{4,\phi\phi}+\tp^2\left(\frac12G_{3,X_1\phi}+G_{4,X_1\phi\phi}\right)
+HA_0f_{5,\phi\phi}-2H\tp(G_{4,X_1\phi}-G_{5,\phi\phi})\right.
\notag\\
&&\hspace{0.9cm}
\left.-H\tp^3(2G_{4,X_1X_1\phi}-G_{5,X_1\phi\phi})
-\frac{H^2\tp^2}{2}\left(3G_{5,X_1\phi}+\tp^2G_{5,X_1X_1\phi}\right)
\right]\dot{H}
\notag\\
&&\hspace{0.9cm}
-\left[\frac12f_{2,X_1\phi}+G_{3,\phi\phi}
+\frac12\tp A_0f_{2,X_1X_2\phi}+\frac{A_0^2}{8}f_{2,X_2X_2\phi}
+\frac12\tp^2(f_{2,X_1X_1\phi}+G_{3,X_1\phi\phi})
-3H\tp(G_{3,X_1\phi}+3G_{4,X_1\phi\phi})\right.
\notag\\
&&\hspace{0.9cm}
\left.
-3H\tp^3\left(\frac12G_{3,X_1X_1\phi}+G_{4,X_1X_1\phi\phi}\right)
+3H^2(G_{4,X_1\phi}-G_{5,\phi\phi})
+3H^2\tp^2(4G_{4,X_1X_1\phi}-\frac52G_{5,X_1\phi\phi})\right.
\notag\\
&&\hspace{0.9cm}
\left.
+3H^2\tp^4\left(G_{4,X_1X_1X_1\phi}-\frac12G_{5,X_1X_1\phi\phi}\right)
+H^3\tp\left(3G_{5,X_1\phi}+\frac72\tp^2G_{5,X_1X_1\phi}
+\frac12\tp^4G_{5,X_1X_1X_1\phi}\right)
\right]\ddot{\phi}
\notag\\
&&\hspace{0.9cm}
-\frac32H^4\tp^2(3G_{5,X_1\phi}+\tp^2G_{5,X_1X_1\phi})
+H^3\left[\frac12A_0(9f_{5,\phi\phi}+A_0^2f_{5,X_3\phi\phi})
-9\tp(G_{4,X_1\phi}-G_{5,\phi\phi})\right.
\notag\\
&&\hspace{0.9cm}
\left.-\tp^3\left(9G_{4,X_1X_1\phi}-\frac72G_{5,X_1\phi\phi}\right)
-\frac12\tp^5G_{5,X_1X_1\phi\phi}\right]
+3H^2\left[2f_{4,\phi\phi}+2G_{4,\phi\phi}+A_0^2f_{4,X_3\phi\phi}\right.
\notag\\
&&\hspace{0.9cm}
\left.+\frac{\dot{A_0}(f_{5,\phi\phi}+A_0^2f_{5,X_3\phi\phi})}{2}
+\tp^2\left(\frac32G_{3,X_1\phi}+3G_{4,X_1\phi\phi}+\frac12G_{5,\phi\phi\phi}\right)
-\tp^4\left(G_{4,X_1X_1\phi\phi}-\frac12G_{5,X_1\phi\phi\phi}\right)\right]
\notag\\
&&\hspace{0.9cm}
-3H\left[\frac{A_0(f_{2,X_2\phi}+4f_{3,\phi\phi})}{4}-A_0\dot{A_0}f_{4,X_3\phi\phi}
+\tp\left(\frac{1}{2}f_{2,X_1\phi}+G_{3,\phi\phi}\right)
-\tp^3\left(\frac12G_{3,X_1\phi\phi}+G_{4,X_1\phi\phi\phi}\right)\right]
\notag\\
&&\hspace{0.9cm}
-\frac14\tp^2(2f_{2,X_1\phi\phi}+\dot{A_0}f_{2,X_1X_2\phi}+2G_{3,\phi\phi\phi})
-\frac{\tp A_0}{4}\left[f_{2,X_2\phi\phi}
+\frac12\dot{A_0}(4f_{2,X_1X_3\phi}+f_{2,X_2X_2\phi})\right]
\notag\\
&&\hspace{0.9cm}
-\dot{A_0}\left(f_{3,\phi\phi}-A_0^2\tilde{f}_{3,\phi\phi}
+\frac{f_{2,X_2\phi}+A_0^2f_{2,X_2X_3\phi}}{4}\right)+\frac{f_{2,\phi\phi}}{2}\,,
\notag\\
&&
D_4=-H^3\tp^2\left(15G_{5,X_1}+10\tp^2G_{5,X_1X_1}+\tp^4G_{5,X_1X_1X_1}\right)
+3H^2\left[A_0(f_{5,\phi}-A_0^2f_{5,X_3\phi})-6\tp(G_{4,X_1}-G_{5,\phi})\right.
\notag\\
&&\hspace{0.9cm}
\left.-\tp^3(12G_{4,X_1X_1}-7G_{5,X_1\phi})-\tp^5(2G_{4,X_1X_1X_1}-G_{5,X_1X_1\phi})\right]
+3H\left[2(f_{4,\phi}+G_{4,\phi})-2A_0^2f_{4,X_3\phi}\right.
\notag\\
&&\hspace{0.9cm}
\left.+\tp^2(3G_{3,X_1}+8G_{4,X_1\phi})+\tp^4(G_{3,X_1X_1}+2G_{4,X_1X_1\phi})\right]
-\tp^3(f_{2,X_1X_1}+G_{3,X_1\phi})-\frac12\tp^2A_0f_{2,X_1X_2}
\notag\\
&&\hspace{0.9cm}
-\tp (f_{2,X_1}-A_0^2f_{2,X_1X_3}+2G_{3,\phi})
+\frac12A_0(f_{2,X_2}+A_0^2f_{2,X_2X_3}+4f_{3,\phi}-4A_0^2\tilde{f}_{3,\phi})\,,
\notag\\
&&
D_5=H^3\left[A_0^3 (f_{5,X_3\phi}+A_0^2f_{5,X_3X_3\phi})
-\tp^3(5G_{5,X_1\phi}+\tp^2G_{5,X_1X_1\phi})\right]
+3H^2\left[2(f_{4,\phi}+G_{4,\phi}+A_0^4f_{4,X_3X_3\phi})\right.
\notag\\
&&\hspace{0.9cm}
\left.+\tp A_0 ( f_{5,\phi\phi}-A_0^2f_{5,X_3\phi\phi} )
-\tp^2(4G_{4,X_1\phi}-3G_{5,\phi\phi})-\tp^4(2G_{4,X_1X_1\phi}-G_{5,X_1\phi\phi})\right]
\notag\\
&&\hspace{0.9cm}
-3H\left[2A_0^3(f_{3,X_3\phi}+\tilde{f}_{3,\phi})
-2\tp(f_{4,\phi\phi}-A_0^2f_{4,X_3\phi\phi}+G_{4,\phi\phi})
-\tp^3(G_{3,X_1\phi}+2G_{4,X_1\phi\phi})\right]
\notag\\
&&\hspace{0.9cm}
-\tp^2(f_{2,X_1\phi}+G_{3,\phi\phi})+2\tp A_0(f_{3,\phi\phi}-A_0^2\tilde{f}_{3,\phi\phi})
+f_{2,\phi}+A_0^2f_{2,X_3\phi}
\,,
\notag\\
&&
D_6=H^2\tp^2(3G_{5,X_1}+\tp^2G_{5,X_1X_1})
-2H\left[A_0f_{5,\phi}-2\tp(G_{4,X_1}-G_{5,\phi})-\tp^3(2G_{4,X_1X_1}-G_{5,X_1\phi})\right]
\notag\\
&&\hspace{0.9cm}
-\tp^2(G_{3,X_1}+2G_{4,X_1\phi})-2(f_{4,\phi}+G_{4,\phi})
\,,
\notag\\
&&
D_7=H^3\tp^2(3G_{5,X_1}+\tp^2G_{5,X_1X_1})
-H^2\left[A_0(3f_{5,\phi}+A_0^2f_{5,X_3\phi})
-6\tp(G_{4,X_1}-G_{5,\phi})-2\tp^3(3G_{4,X_1X_1}-2G_{5,X_1\phi})\right]
\notag\\
&&\hspace{0.9cm}
-H\left[2(f_{4,\phi}+2A_0^2f_{4,X_3\phi}+G_{4,\phi})-2A_0\tp f_{5,\phi\phi}
+\tp^2(3G_{3,X_1}+10G_{4,X_1\phi}-2G_{5,\phi\phi})\right]
\notag\\
&&\hspace{0.9cm}
+\tp(f_{2,X_1}+2f_{4,\phi\phi}+2G_{3,\phi}+2G_{4,\phi\phi})
+\frac12A_0(f_{2,X_2}+4f_{3,\phi})
\,,
\notag\\
&&
D_8=-\frac{2\tp D_1+D_4+3HD_6}{A_0}\,,
\notag\\
&&
D_9=-H^3A_0^2(3f_{5,X_3\phi}+A_0^2f_{5,X_3X_3\phi})
-3H^2\left[2A_0(f_{4,X_3\phi}+A_0^2f_{4,X_3X_3\phi})
-\tp(f_{5,\phi\phi}+A_0^2f_{5,X_3\phi\phi})\right]
\notag\\
&&\hspace{0.9cm}
+6HA_0\left[A_0(f_{3,X_3\phi}+\tilde{f}_{3,\phi})+\tp f_{4,X_3\phi\phi}\right]
-\tp\left(\frac12f_{2,X_2\phi}+2f_{3,\phi\phi}-2A_0^2\tilde{f}_{3,\phi\phi}\right)
-A_0f_{2,X_3\phi}
\,,
\notag\\
&&
D_{10}=-2\dot{H}f_{5,\phi}-H^2 \left(3f_{5,\phi}+A_0^2f_{5,X_3\phi} \right)
-2HA_0 \left(2f_{4,X_3\phi}+\dot{A_0}f_{5,X_3\phi} \right)
-2\dot{A_0}f_{4,X_3\phi}+2f_{3,\phi}+\frac{1}{2} f_{2,X_2}\,,
\ea
and 
\ba
&&
w_1=-H^2\left[A_0^3(f_{5,X_3}+A_0^2f_{5,X_3X_3})-\tp^3(5G_{5,X_1}+\tp^2G_{5,X_1X_1})\right]
-2H\left[2(f_4+A_0^4f_{4,X_3X_3}+G_4)\right.
\notag\\
&&\hspace{0.9cm}
\left.+A_0\tp(f_{5,\phi}-A_0^2f_{5,X_3\phi})-\tp^2(4G_{4,X_1}-3G_{5,\phi})
-\tp^4(2G_{4,X_1X_1}-G_{5,X_1\phi})\right]
\notag\\
&&\hspace{0.9cm}
-\tp^3(G_{3,X_1}+2G_{4,X_1\phi})-2\tp(f_{4,\phi}-A_0^2f_{4,X_3\phi}
+G_{4,\phi})+2A_0^3(f_{3,X_3}+\tilde{f}_3)
\,,
\notag\\
&&
w_2=w_1+2Hq_t-\tp D_6\,,
\notag\\
&&\hspace{0.51cm}
=A_0 \left[
-H^2A_0^2(3f_{5,X_3}+A_0^2f_{5,X_3X_3})
-2H\left[2A_0(f_{4,X_3}+A_0^2f_{4,X_3X_3})
-\tp(f_{5,\phi}+A_0^2f_{5,X_3\phi})\right]
+2A_0\tp f_{4,X_3\phi}
\right.\notag\\
&&\hspace{0.8cm}
\left.
+2A_0^2( f_{3,X_3}+\tilde{f}_3) \right]\,,
\notag\\
&&
w_3=-2A_0^2q_v\,,
\notag\\
&&
w_4=w_5-H^3\left[3A_0^3(2f_{5,X_3}+A_0^2f_{5,X_3X_3})
-\tp^3\left(15G_{5,X_1}+\frac{13}{2}\tp^2G_{5,X_1X_1}+\frac12\tp^4G_{5,X_1X_1X_1}\right)\right]
-3H^2\left[2(f_4+G_4)\right.
\notag\\
&&\hspace{0.9cm}
\left.+A_0^2\left(2f_{4,X_3}+4A_0^2f_{4,X_3X_3}-3A_0\tp f_{5,X_3\phi}\right)
-\tp^2(7G_{4,X_1}-6G_{5,\phi})-\tp^4\left(8G_{4,X_1X_1}-\frac92G_{5,X_1\phi}\right)\right.
\notag\\
&&\hspace{0.9cm}
\left.-\tp^6\left(G_{4,X_1X_1X_1}-\frac12G_{5,X_1X_1\phi}\right)\right]
+3H\left[2A_0^3(f_{3,X_3}+\tilde{f}_3)-2\tp(f_{4,\phi}-2A_0^2f_{4,X_3\phi}+G_{4,\phi})\right.
\notag\\
&&\hspace{0.9cm}
\left.-\tp^3(2G_{3,X_1}+5G_{4,X_1\phi})-\tp^5\left(\frac12G_{3,X_1X_1}+G_{4,X_1X_1\phi}\right)\right]
+\frac12\tp^4(f_{2,X_1X_1}+G_{3,X_1\phi})
\notag\\
&&\hspace{0.9cm}
+\tp^2\left(\frac12f_{2,X_1}-A_0^2f_{2,X_1X_3}-\frac18A_0^2f_{2,X_2X_2}+G_{3,\phi}\right)
-\frac12A_0\tp\left(f_{2,X_2}+A_0^2f_{2,X_2X_3}+4f_{3,\phi}-4A_0^2\tilde{f}_{3,\phi}\right)
\,,
\notag\\
&&
w_5=\frac12H^3A_0^3(3f_{5,X_3}+6A_0^2f_{5,X_3X_3}+A_0^4f_{5,X_3X_3X_3})
+3H^2A_0\left[A_0^3(3f_{4,X_3X_3}+A_0^2f_{4,X_3X_3X_3})\right.
\notag\\
&&\hspace{0.9cm}
\left.+\frac12\tp(f_{5,\phi}-2A_0^2f_{5,X_3\phi}-A_0^4f_{5,X_3X_3\phi})\right]
-3HA_0^3\left[f_{3,X_3}+\tilde{f}_3+A_0^2(f_{3,X_3X_3}+\tilde{f}_{3,X_3})
+A_0\tp f_{4,X_3X_3\phi}\right]
\notag\\
&&\hspace{0.9cm}
+\frac18A_0^2\tp^2f_{2,X_2X_2}-\frac14A_0\tp\left[f_{2,X_2}+4f_{3,\phi}
-2A_0^2(f_{2,X_2X_3}-2\tilde{f}_{3,\phi}+2f_{3,X_3\phi})+4A_0^4\tilde{f}_{3,X_3\phi}\right]
+\frac12A_0^4f_{2,X_3X_3}
\,,
\notag\\
&&
w_6=-\frac{w_1-\tp D_6+2Hq_t}{A_0}-4H\left(2A_0f_{4,X_3}-\tp f_{5,\phi}
+HA_0^2f_{5,X_3}\right)\,,
\notag\\
&&\hspace{0.51cm}
=-H^2A_0^2(f_{5,X_3}-A_0^2f_{5,X_3X_3})
-2H\left[2A_0(f_{4,X_3}-A_0^2f_{4,X_3X_3})
-\tp (f_{5,\phi}-A_0^2f_{5,X_3\phi})\right]
-2 A_0\tp f_{4,X_3\phi}\notag\\
&&\hspace{0.9cm}
-2A_0^2( f_{3,X_3}+\tilde{f}_3)\,,
\notag\\
&&
w_7=-2\dot{H} \left(2f_{4,X_3}+HA_0f_{5,X_3} \right)
-H^2\left[\frac{\tp(3f_{5,\phi}+A_0^2f_{5,X_3\phi})}{A_0}
+\dot{A_0} \left(f_{5,X_3}+A_0^2f_{5,X_3X_3} \right)\right]\notag\\
&&\hspace{0.9cm}
-4H \left( \tp f_{4,X_3\phi}+A_0\dot{A_0}f_{4,X_3X_3} \right)
+2\dot{A_0} \left( f_{3,X_3}+\tilde{f}_3 \right)
+\frac{\tp( f_{2,X_2}+4f_{3,\phi})}{2A_0}\,,
\notag\\
&&
w_8=3Hw_1-2w_4-\tp D_4\,.
\ea
We note that we used other background 
Eqs.~(\ref{back1}) and (\ref{back4}) for the derivation 
of these coefficients.


\end{document}